# How do academic topics shift across altmetric sources? A case study of the research area of Big Data

Xiaozan Lyu[1, 2]

lvxz1991@zju.edu.cn

[1] Department of Information Resources Management, School of Public Affairs, Zhejiang University, Hangzhou 310058, China

[2] Center for Science and Technology Studies (CWTS), Leiden University, Kolffpad 1, P.O. Box 905, 2300 AX Leiden, The Netherlands

Rodrigo Costas[2, 3]

rcostas@cwts.leidenuniv.nl

[2] Center for Science and Technology Studies (CWTS), Leiden University, Kolffpad 1, P.O. Box 905, 2300 AX Leiden, The Netherlands

[3] Centre for Research on Evaluation, Science and Technology (CREST), Stellenbosch University, RW Wilcocks Building, Stellenbosch, 7600, Stellenbosch, South Africa

## Abstract

Taking the research area of Big Data as a case study, we propose an approach for exploring how academic topics shift through the interactions among audiences across different altmetric sources. Data used is obtained from Web of Science (WoS) and Altmetric.com, with a focus on Blog, News, Policy, Wikipedia, and Twitter. Author keywords from publications and terms from online events are extracted as the main topics of the publications and the online discussion of their audiences at Altmetric. Different measures are applied to determine the (dis)similarities between the topics put forward by the publication authors and those by the online audiences. Results show that overall there are substantial differences between the two sets of topics around Big Data scientific research. The main exception is Twitter, where high-frequency hashtags in tweets have a stronger concordance with the author keywords in publications. Among the online communities, Blogs and News show a strong similarity in the terms commonly used, while Policy documents and Wikipedia articles exhibit the strongest dissimilarity in considering and interpreting Big Data related research. Specifically, the audiences not only focus on more easy-to-understand academic topics related to social or general issues, but also extend them to a broader range of topics in their online discussions. This study lays the foundations for further investigations about the role of online audiences in the transformation of academic topics across altmetric sources, and the degree of concern and reception of scholarly contents by online communities.

**Keywords:** Altmetrics; Topic shift; Similarity measurements; Big Data research area

# Introduction

Social media has been recognized as "the most pervasive form of communication in all fields today" (McCaughey et al. 2014), profoundly changing the way people interact with one another. Social media is also influencing and changing the way how science and academic topics are being communicated nowadays (Sugimoto et al. 2017). According to the estimation of the company Altmetric.com, around 15,000 unique research outputs are shared or mentioned online each day and a research output is mentioned online every 1.8 seconds (Altmetric 2016). Some scholars (e.g., Rowlands et al. 2011; Van Noorden 2014; Haustein 2016) argue that social media can promote openness and transparency, making the process of peer-review more visible, and with scholarly ideas and results being more openly discussed and scrutinized in the social media realm. In addition, social media attention to scholarly research can help increase the public attention to science. The academic social media users (especially the researchers) can quickly disseminate their studies and publications, *pushing* knowledge to their audiences straightly (Allen et al. 2013).

The transformative power of social media in scholarly communication, opens up a way for the study of social media impact (i.e. popularity, attention, visibility, etc.) of scientific research, making it a whole new research area in the field of Scientometrics (Bornmann 2014; Bornmann and Haunschild 2017). The analysis and study of the interactions between social media and scholarly agents and products (Haustein et al. 2016), popularly known as "altmetrics" and more specifically as "Social Media Metrics" (SMM) of science, have opened a new analytical scientometric perspective, with the potential to complement the more traditional citation-based indicators, expanding the understanding of how scientific ideas and topics are discussed and disseminated across multiple diverse communities (Costas 2018).

An important characteristic of SMM of science is their large source and metric heterogeneity. This heterogeneity goes from studies of the mentions to scientific articles on microblogging platforms like Twitter and Weibo, to posts about scientific research on social network sites such as Facebook and Google+, saves of scientific references on online reference managers like Mendeley and CiteULike, reviews on F1000Prime, Publons or PubPeer, as well as mentions in scholarly blogs, news and mainstream media (e.g., Haunschild and Bornmann 2015*;* Haustein, Costas and Larivière 2015; Thelwall 2017; Maflahi and Thelwall 2018; Robinson-Garcia et al. 2019). Previous research in the field have also focused on studying the most important sources providing altmetric data (e.g., Thelwall et al. 2013; Wouters and Costas 2012; Zahedi, Costas and Wouters 2014), the coverage of scientific publications across altmetric sources (i.e. the percentage of documents with at least one mention on a particular social media platform) (e.g., Alperin 2015; Costas, Zahedi and Wouters 2015; Haustein, Costas and Larivière 2015), and the correlation between these new metrics and the traditional bibliometric indicators as well, particularly with citation impact (e.g., Costas, Zahedi and Wouters 2015; Haustein et al. 2014; Thelwall et al. 2013).

In addition to the role of social media in increasing the visibility of scholars and their work, research around SMM of science have also attempted to trace the public perceptions and opinions from online communities about specific scientific fields or topics, for instance,

"climate change" (e.g., An et al. 2014; Pearce et al. 2014; Haustein et al. 2014), "Rio+20[1]" (Hellsten and Leydesdorff 2017), and "migrant crisis" (Nerghes and Lee 2018). In a recent study, Haunschild and his colleagues (2019) explored a novel network approach to compare topics between researchers and Twitter users based on author keywords and Twitter hashtags, offering insights that publications being tweeted can clearly be distinguished from those that are not tweeted. This type of studies put the emphasis in the "inherently social" nature characteristic of the altmetric sources like Blog or Twitter (Walker 2006), where the forwarding and commenting functionalities make it possible for "the shift from public understanding to public engagement with science" (Kouper 2010; Sugimoto et al. 2017).

As highlighted by Sugimoto and her colleagues (2017), the broader social impacts should not be conceived merely as a distinction of the audiences who receive the work, or as a recognition of the work that catches the attention of audiences, but rather as the amplification of different voices which are disseminating and attracting the attention. In fact, social media is more than just marketing for academic work. It can inform every step of the research process: helping researchers get a pulse on the different movements in the fields or topics they are interested in, assisting in the promotion of published work, and also contributing to harvest helpful feedback for further research (Alampi 2012).

Accordingly, we argue that in addition to focusing on the potential alternative role of social media in assessing research impact, exploring their role in the dynamics and patterns of cross-platform or cross-community shift of academic topics is also of great value. This paper will contribute to this aim. Taking the research area of Big Data as a case study, we attempt to investigate the semantic similarity between topics from publications and those from the discussions of audiences mentioning and disseminating publications across different altmetric sources, including Blogs, News, Policy documents[2], Wikipedia and Twitter. To be more specific, we want to answer the following questions:

1) What are the most important *academic topics* represented by the high-frequency author keywords in Big Data publications?
2) How do online audiences from different altmetric sources deal with the academic topics in their online discussions? In essence, how (dis)similar are the terms used by both communities (academic and online) in representing the same publications?
3) More specifically, on which platform are the audiences' terms more consistent with those of Big Data publications (i.e. author keywords)? And, in the online community, on which platforms do the online audiences use more similar terms in their discussions?

# Methodology

We used the Web of Science (WoS) and altmetric data from the Centre for Science and Technology Studies (CWTS) in-house databases derived from the Science Citation Index

[1] "Rio+20" refers to the United Nations Conference on Sustainable Development, which was held in Rio de Janeiro, Brazil on 20 to 22 June 2012. https://www.environment.gov.au/about-us/international/rio-20

[2] In our view, Policy document mentions may not be seen as strictly social media events (see also Wouters, Zahedi and Costas 2018); however we decided still to include them in this study as a relevant source by itself in capturing forms of policy-related impact (Bornmann, Haunschild and Marx 2016).

Expanded (SCI-E), Social Sciences Citation Index (SSCI), and Arts and Humanities Citation Index (AHCI), as well as Altmetric.com[3]. A comprehensive list of 9,596 scholarly documents (i.e. Article, Review, and Letter) related with the research area of Big Data was obtained (we refer to them as *Big Data publications*) by using the search terms "big data" or "bigdata" in title, abstract and keywords of publications. About 90% (that is 8,626) of all the publications have a Digital Object Identifier (DOI) in the WoS database, which enable us to match these publications with the altmetric data. Although not all publications related to the research area of Big Data can be covered with our search strategy, such a narrow but precise approach is the most efficient in terms of unambiguously identifying publications that have the most unambiguous alignment with the core concept of "Big Data".

From a social media metric point of view, once a publication is mentioned in a post on an altmetric platform, a publication-post linkage is established. The online user who published this post can be seen as the *audience* of the publication mentioned. We propose a conceptual model of the process of topic spreading from academia to different altmetric sources (see Fig. 1).

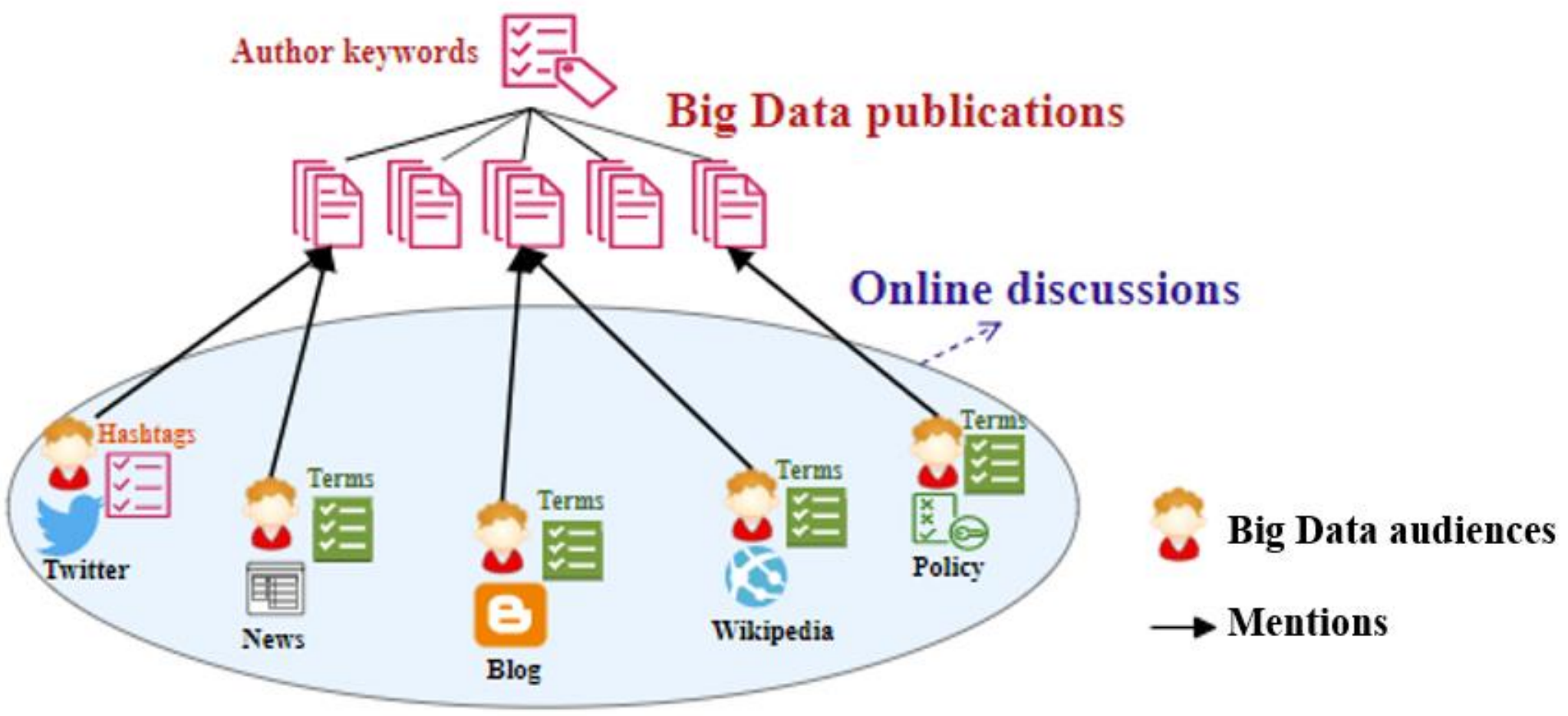

**Fig. 1** Instance of topic spreading model across altmetric sources.

In this model, the online audiences from the five platforms (i.e. Twitter, Blog, Policy, News, and Wikipedia) mentioning at least one Big Data publication are the *Big Data audiences* of the publications. That is to say, these audiences wrote and posted online events referencing these publications, which constitute the *online discussions* (the blue circle) about the research area of Big Data. In this way, the online events can be seen as a channel, through which the academic topics are spread and potentially amplified from the academic community to the online community. In order to further explore the topic similarity between the two communities, the author keywords from the publications and textual terms from the online events were extracted and processed. Technically, concerning the differences in text structure, title or summary terms of blogs, news, policy documents, and Wikipedia articles, and hashtags of tweets, are extracted separately, which also divides the online audiences into two groups. The concepts in the model are detailed as follows:

[3] https://www.altmetric.com. The data from Altmetric.com used in this study is updated up to October 2017.

- *Big Data publications*: scientific publications included in our dataset directly using “big data” in title, abstract and author keywords. The authors of these Big Data publications are simply referred to as *Big Data authors*.
- *Big Data audiences*: users across the five platforms (Twitter, Blog, Policy, News, and Wikipedia) who have mentioned at least one Big Data publication, and are further divided into two groups:
    - *Audiences on Twitter*
    - *Audiences on Blog, News, Policy and Wikipedia*[4]
- *Big Data topics:* high-frequency *author keywords (K)* from publications and terms from social media events. Specifically, two approaches are applied to acquire the terms from the two audience groups:
    - *Text terms (T):* terms generated from titles of blogs, news and policy documents, as well as the first sentence in summaries of Wikipedia articles.
    - *Hashtags (H)*: terms starting with the # sign from tweets, which is a system of categorization within Twitter and has a similar function of the author keywords in publications (Haustein 2016)[5].

Of the 8,626 publications with DOIs, 3,563 (41.3%) have been mentioned at least once on any of the five altmetric sources, of which 3,493 (40.5%) have been tweeted by Twitter users, 697 (8.1%) by users from any of the other four platforms, and 627 (7.3%) by audiences in both of the two groups (Table 1).

**Table 1** Statistic description of data used in the study

| | **Group 1** | | | | **Group 2** |
|---|---|---|---|---|---|
| | **News** | **Policy** | **Blog** | **Wikipedia** | **Twitter** |
| N(all events) | 2,825 | 111 | 1,105 | 179 | 74,450 |
| N(unique events) | 1,855 | 90 | 974 | 146 | 42,341 |
| N(mentioned papers) | 367 | 85 | 412 | 125 | 3,493 |
| | 697 | | | | |
| Share (in 8,626, %) | 4.25 | 0.99 | 4.78 | 1.45 | 40.5 |
| | 8.1 | | | | |
| **Grant Total** | **3,563 (41.3%)** | | | | |

According to the model, we divide our research process into several steps:

**1) Identification of topics of publications and online audiences.** VOSviewer (Van Eck and Waltman 2010) was used for extracting high-frequency author keywords, hashtags and textual terms as topics of the three groups, respectively. Considering the differences in the numbers of topics in each group, we uniformly selected the top-100 topics with the highest frequency. The text mining functionality of VOSviewer provides support for creating term maps based on a corpus of documents with the following steps (Van Eck & Waltman, 2011):

[4] Considering the short titles of Wikipedia articles, we choose to use the first sentence in the summary which is a condensed explanation of an event, and is equivalent to the titles of blogs, news and policy documents in part.
[5] This decision is also backed up by the results observed by Robinson-Garcia et al (2017) in which they found relatively low levels of engagement of tweeters with publication, therefore limiting the value of a semantic study based only on tweets’ full text.

i. Identification of noun phrases with an approach developed by Van Eck, Waltman, Noyons and Buter (2010). The linguistic filter which selects all word sequences that consist exclusively of nouns and adjectives and that end with a noun was used to identify noun phrases.
ii. Selection of the most relevant noun phrases. The selected noun phrases are referred to as terms. For each noun phrase, the distribution of (second-order) co-occurrences over all noun phrases is determined. The larger the difference between the two distributions, the higher the relevance of a noun phrase. Then, noun phrases with a high relevance are grouped together into clusters.
iii. Mapping and clustering of the terms. The unified framework for mapping and clustering (Van Eck, Waltman, Dekker, & Van den Berg, 2010; Waltman, Van Eck, & Noyons, 2010) is used in this step.
iv. Visualization of the mapping and clustering results.

**2) Similarity measurement.** Cosine similarity measurement was applied to quantitatively investigate the degree of (dis)similarity among topic sets of different groups, and is formulated as follows:

$$\text{Similarity} = \frac{A \cdot B}{||A||\,||B||} = \frac{\sum_{i=1}^{n} A_i B_i}{\sqrt{\sum_{i=1}^{n} A_i^2}\sqrt{\sum_{i=1}^{n} B_i^2}} \qquad (1)$$

In Eq. (1), $A_i$ and $B_i$ are components of vector A and B, respectively (different topic sets in our study). The resulting similarity ranges from −1 meaning exactly opposite, to 1 meaning precisely the same, with 0 indicating orthogonality or decorrelation, while in-between values indicate intermediate similarity or dissimilarity (Huang 2008).

**3) Comparison of different types of topics.** All the topics can be classified into four non-overlapping types on the basis of their occurrences in groups:

- KTH: topics that appear in all groups as author keywords, terms, and hashtags, which can be considered as the *common topics* of both publications and online audiences;
- K: topics that appear only as author keywords, and can be considered as the *pure academic topics*;
- T/H/TH: topics that appear only as terms and hashtags, which can be regarded as the *pure audience topics*, alternatively, one can say that they are to some extent the *amplification of academic topics*[6] in online communities;
- KT/KH: topics that appear in author keywords and any other group of terms (i.e. hashtags or text terms).

The analysis of the different types of topics helps to comprehend and interpret the tendency of focus of publications and online audiences around the research area of Big Data, as well as the pattern of how the topics shift from academia to the online community.

[6] We can argue that these topics are added by the online users, thus "expanding" or "amplifying" the initial topics put forward by the authors through the author keywords. It could also be argued, that these topics added by the online users are also a sort of "reinterpretation" of the academic topics of the papers.

# Results

A number of different analyses are performed in order to answer the research questions stated above. This section presents the results of these analyses, including topic identification, similarity analysis, and comparison among topics of groups.

## Identification of topics

*Author keywords*

Of all the 8,626 publications, 6,689 (about 78%) have a total of 19,065 author keywords with a sum of 36,362 occurrences in total. The top-100 author keywords as the topics of Big Data publications account for approximately 22.6% over all the occurrences. Fig. 2 shows the cluster map[7] of these author keywords based on their co-occurrences in Big Data publications. Each item represents an author keyword. The size of an item indicates the number of total occurrences of the corresponding item. The color of an item represents the main cluster to which it belongs. The distance between two items offers an approximate indication of the relatedness in terms of their co-occurrences.

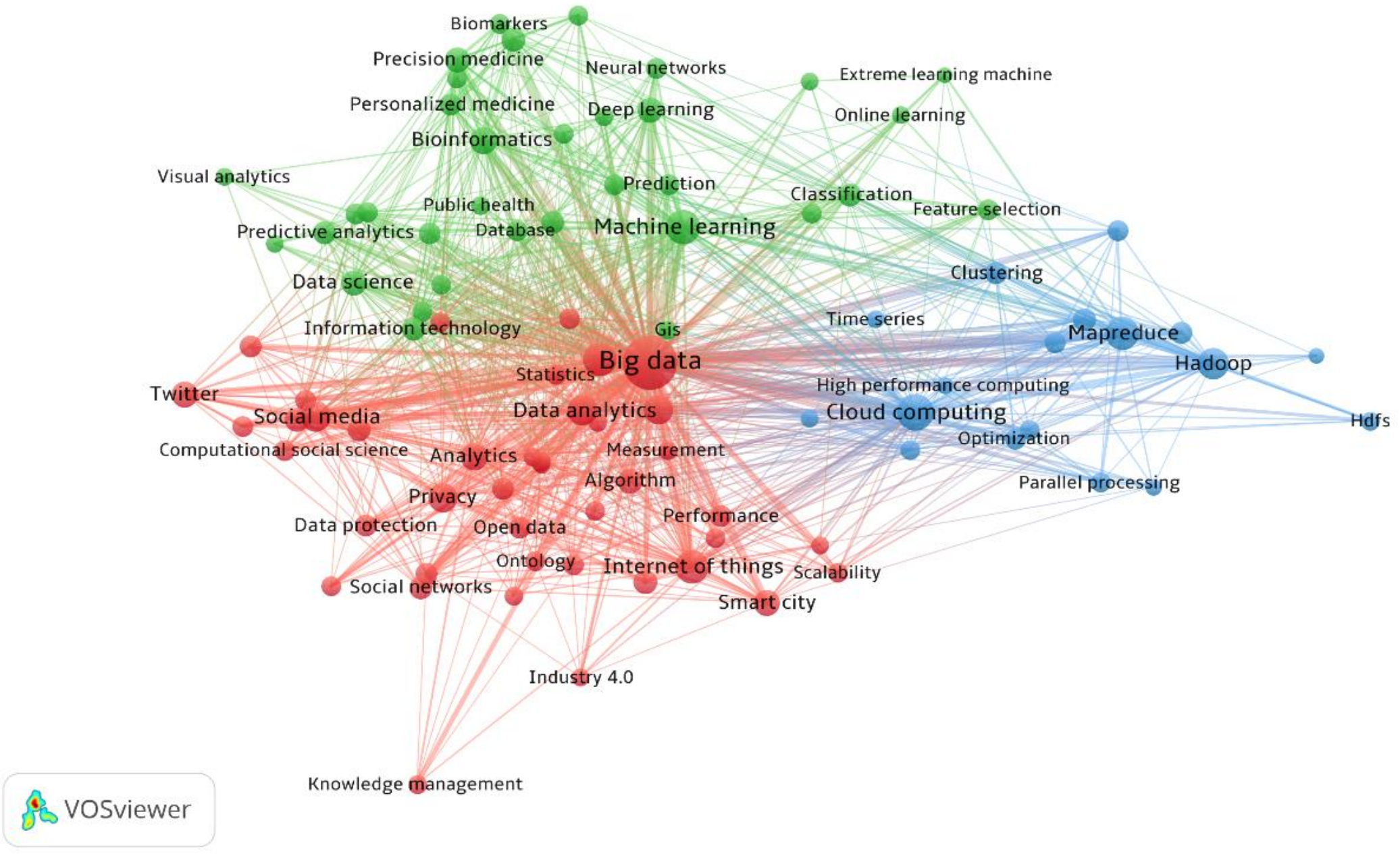

**Fig. 2** Cluster map of high-frequency author keywords of publications.

This term map provides us a clear overview of the main author keywords of the Big Data publications. Three different clusters can be identified. The red cluster is the largest group containing the most author keywords (i.e. 44), of which many are related to social issues from industrial development to social media, such as "Internet of things", "social media", and

[7] VOSviewer is used for clustering author keywords: a resolution of 0.5 is employed in the clustering algorithm, with minimal cluster size of 1 item, and the option "merge small clusters" is enabled. The "association strength" is applied for normalization. Default values are used for layout.

"Industry 4.0". The green cluster contains terms of the applications of data analytical technologies in bioscience and medicine, for instance, "Bioinformatics" and "Precision medicine". This is the second largest cluster consisting of 31 terms. The blue cluster, is the smallest one, it is mainly focused on core technologies with technical terms, especially machine learning and cloud computing-related techniques (e.g., "Cloud computing", "Hadoop", and "Mapreduce")[8]. Although the keyword "Machine learning" locates in the green cluster, it is quite close to the technology cluster. It follows that the top-100 author keywords seem to cover from core technologies of Big Data to major applications and social impact.

Table 2 details the top-10 author keywords with the highest frequency. It is remarkable that the search term "Big data" only appears in 44% of all the publications as an author keyword, indicating that instead of tagging their publications with this term as an author keyword straightly, most Big Data publications just mentioned it in title or abstract. The second to fifth places on the list are all technology-related terms (i.e. "Cloud computing", "Machine learning", "Data mining", and "Mapreduce"). However, these four topics only appear in about 3.5% of publications on average, demonstrating the diverse and scattered topicality around the research area of Big Data. The high frequency of "Social media", "Internet of Things", and "Privacy" implies that, the opportunities and challenges brought by the explosion of massive data have aroused great concern and discussion among scholars, especially those in the social sciences.

**Table 2** Top-10 high-frequency author keywords

| | **Keywords** | **Occurrences** | **Share (%)** | **Share in publications (%)** | **Cluster** |
|---|---|---|---|---|---|
| 1 | Big data | 2,960 | 8.14 | 44.25 | 1 |
| 2 | Cloud computing | 328 | 0.90 | 4.90 | 2 |
| 3 | Machine learning | 264 | 0.73 | 3.95 | 3 |
| 4 | Data mining | 244 | 0.67 | 3.65 | 1 |
| 5 | Mapreduce | 232 | 0.64 | 3.47 | 2 |
| 6 | Social media | 163 | 0.45 | 2.44 | 1 |
| 7 | Big data analytics | 159 | 0.44 | 2.38 | 1 |
| 8 | Hadoop | 139 | 0.38 | 2.08 | 2 |
| 9 | Internet of things | 126 | 0.35 | 1.88 | 1 |
| 10 | Privacy | 107 | 0.29 | 1.60 | 1 |

*Title or summary terms*

A total of 3,063 titles or summaries of posts mentioning Big Data publications in blogs, news, policy documents, and Wikipedia articles, are obtained. Among all the items, 1,855 (60.6%) are from Blogs, 973 (32.1%) are News titles, while Wikipedia and Policy only account for 4.8% (146 summaries) and 2.9% (89 titles) respectively. Altogether, 5,512 terms with 9,447 occurrences are extracted by VOSviewer with the same approach as we did for author keywords. Fig. 3 shows the map of the top-100 high-frequency terms divided into four clusters.

The largest cluster containing almost half (45, red) of all the terms is related to general issues, typically of medical science and health care (e.g., "Patient", "Mental health", "Disease", and "Depression"). In addition, some social media related events like "Tweet", "App" and

[8] Map-reduce and Hadoop are the two leading tools related with machine learning and cloud computing (Zhang et al. 2019).

"Instagram" also have received a lot of attention. The green one covers terms associated with scientists and research, for instance, "Scientist", "Study" and "Publication", and is the second largest with a total of 42 terms. Terms about interpersonal relationships and political affairs are distributed across the other two smaller clusters (i.e. blue and yellow). Besides, "Facebook" has the most links in the network, far more than "Big data", illustrating its popularity among the online audiences. Nonetheless, due to the skewed distribution of links, "Facebook" is the center of the cluster it belongs to, but not the center of the whole network.

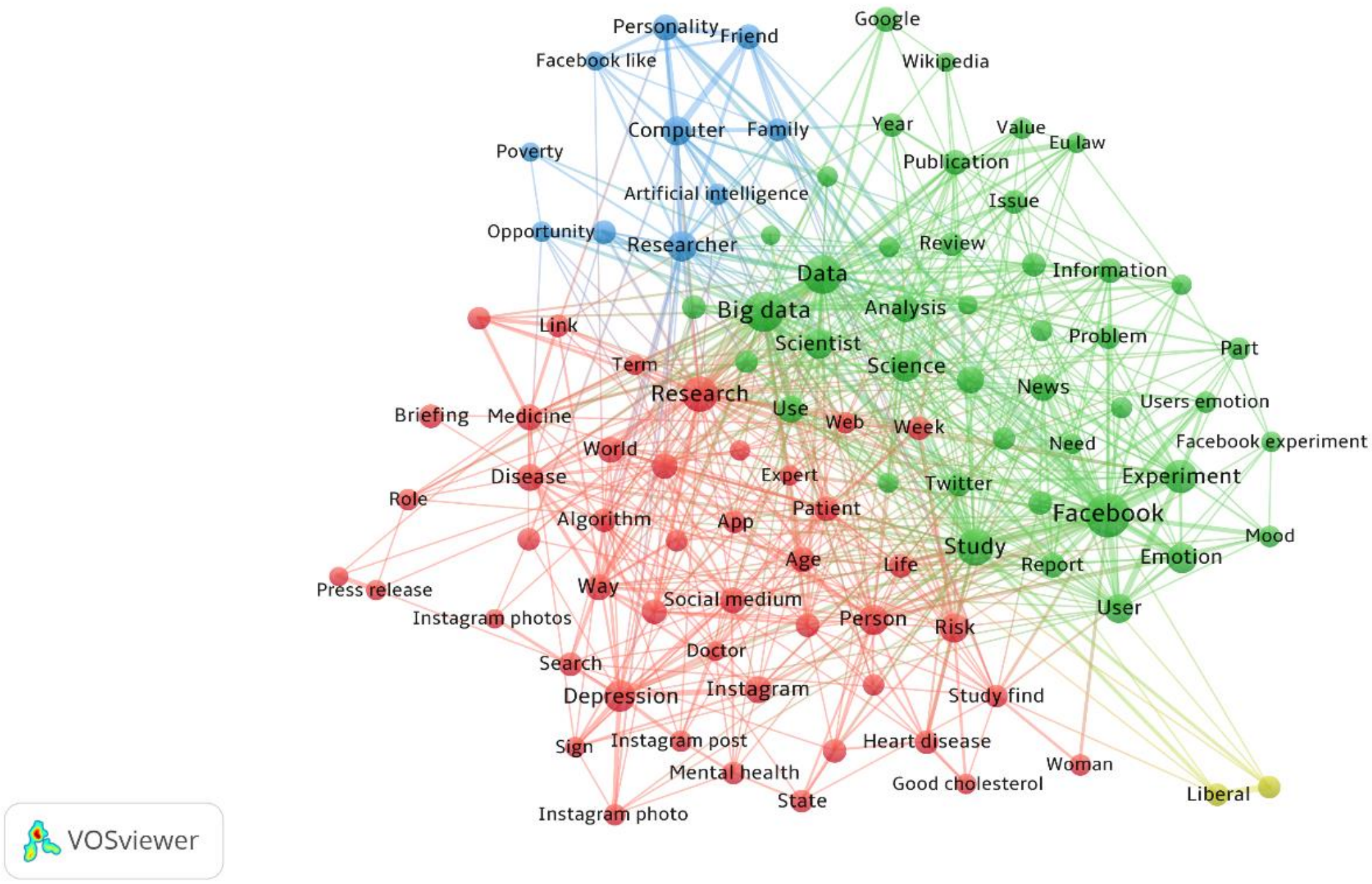

**Fig. 3** Cluster map of high-frequency terms from Blog, News, Policy, and Wikipedia.

"Facebook" ranks first among the top-10 high-frequency terms, appearing in 273 (8.94%) entries in all, surpassing "Big Data" ranking second (184, 5.89%). It may signal to some extent the shift in the focus of the online community around Big Data publications, compared to the focus among the academic scholars. The high frequency of "Study", "Research" and "Science" highlights the importance of scientific literature as a main information source of these posts. In addition, mental health-related terms, like "Depression" and "Emotion", also have gained substantial attention from the audiences, which is one of the main application fields of Big Data analysis technologies closely related to individuals (Table 3).

**Table 3** Top-10 high-frequency text terms

| | Terms | Occurrences | Share (%) | Share of events (%) | Cluster |
|---|---|---|---|---|---|
| 1 | Facebook | 275 | 3.01 | 8.94 | 2 |
| 2 | Big data | 184 | 2.01 | 5.98 | 2 |
| 3 | Data | 139 | 0.49 | 4.52 | 2 |
| 4 | Study | 125 | 0.64 | 4.07 | 2 |
| 5 | Research | 104 | 0.39 | 3.38 | 1 |
| 6 | Experiment | 85 | 0.20 | 2.76 | 2 |
| 7 | Depression | 66 | 0.57 | 2.15 | 1 |
| 8 | Science | 59 | 0.27 | 1.92 | 2 |
| 9 | Emotion | 53 | 0.34 | 1.72 | 2 |
| 10 | Researcher | 52 | 0.44 | 1.69 | 3 |

The overlay maps in Fig. 4 further display the sources of these terms, as well as their occurrences on each platform. The overlay scores used in these maps are normalized by dividing by the mean, so that the four sources can be compared with each other. The color depth of a term is based on its overlay score. That is to say, the higher the frequency, the darker the color. The gray term means that it does not appear in the corresponding source.

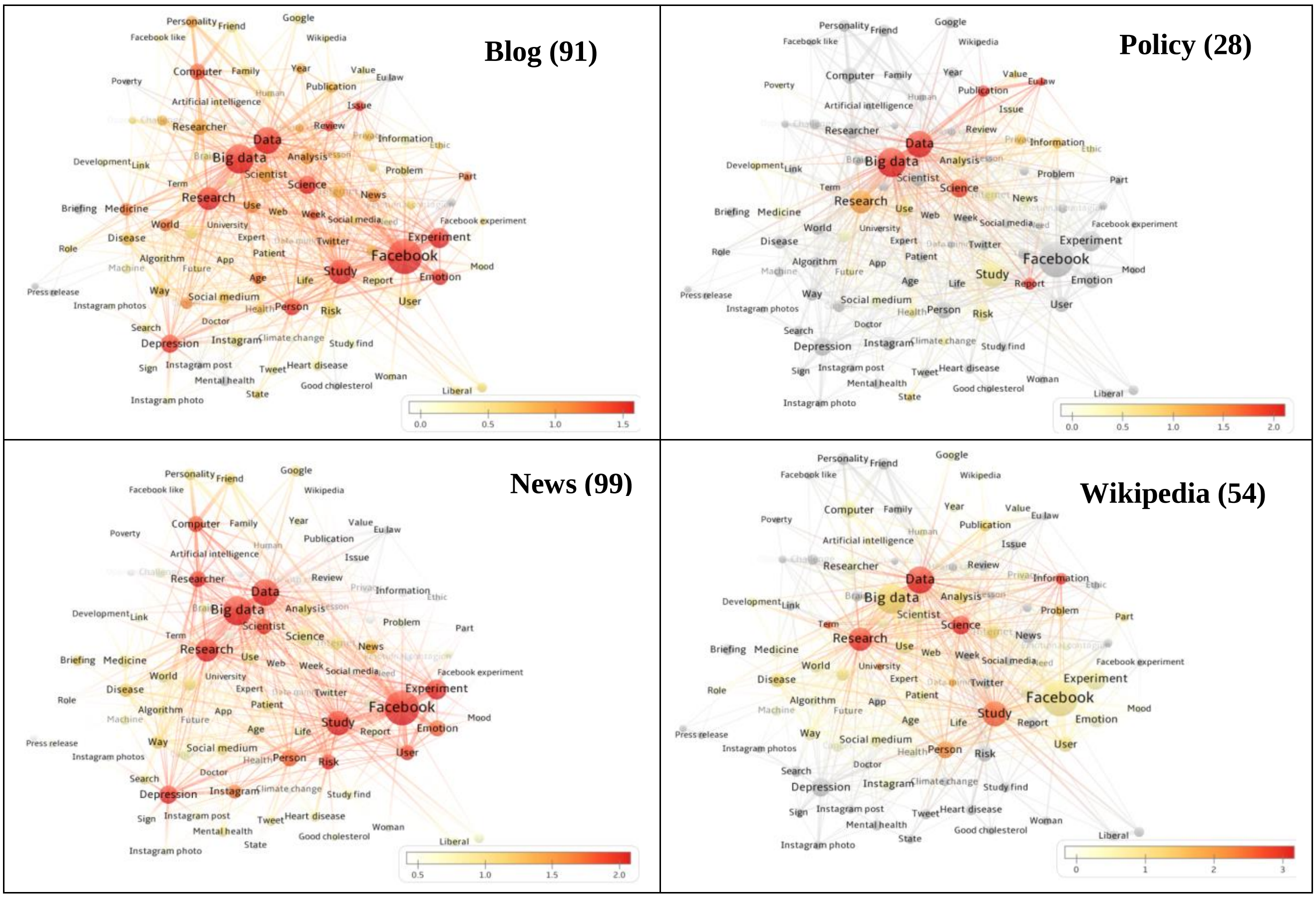

**Fig. 4** Overlay maps of terms from Blog, Wikipedia, News, and Policy. In brackets is the number of terms finally selected.

It is revealing that blogs and news contribute more terms due to their larger numbers of involved titles, among which topics related to social media, health care, and science are the common interest of the users on these two platforms (e.g., “Facebook”, “Emotion” and “Research”). Besides, news have a more extensive range of focus than blogs, covering more diverse terms ranging from medicine (e.g., “Mental health” and “Alzheimer”), to technologies and some social issues (e.g., “Nanotechnology” and “Poverty). By comparison, policy documents and Wikipedia entries have a more limited focus on Big Data publications with fewer publications mentioned. Specifically, the high-frequency terms in these two groups suggest a quite a different concern of topics on these platforms. Wikipedia entries are more oriented towards the research and application of technologies on internet and web, while policy documents have an obvious orientation to more general issues related to social progress like “EU law” and “Climate change”.

In the *Appendix,* we also provide four cluster maps of terms extracted from titles of blogs, news, policy documents, and first sentences of summaries of Wikipedia articles, separately (Fig. 13). Because of the quantity variance of entities, the minimum number of occurrence for being plotted is 3 for terms from Blogs and News, and 2 for terms from Policy documents and Wikipedia articles. The results shown in these figures differ rarely from those obtained by the approach described above. Blogs and News media mentioning Big Data publications have a stronger semantic relationship with topics around medicine, health care, social media research, and technologies. Policy documents citing Big Data publications tend to focus more on political, legal or social issues related with Big Data (e.g., "eu law", "privacy", or "policy"), while mentions of Big Data publications from Wikipedia are more oriented towards academic, technical and more theoretical topics (e.g., "university", "cloud computing", or "theory").

*Hashtags of tweets*

A total of 4,566 hashtags from 42,341 distinct tweets are obtained. These hashtags have a sum of 41,412 occurrences in all. Like other groups, the cluster map is provided in Fig. 5 with four clusters integrated by the top-100 high-frequency hashtags. The red cluster contains various terms related to bioscience and medicine, such as "#Genomics", "#Genetics", "#Cancer", "#Bioinformatics" and "#Precisionmedicine". The green one covers not only core technologies like "#Machinelearning" and "#AI", but also terms about health care (e.g., "#Healthit" or "#Digitalhealth"). The blue cluster contains topics mostly related to social media and social networks, typically as "#Facebook". The yellow cluster is focused on economic development and social management.

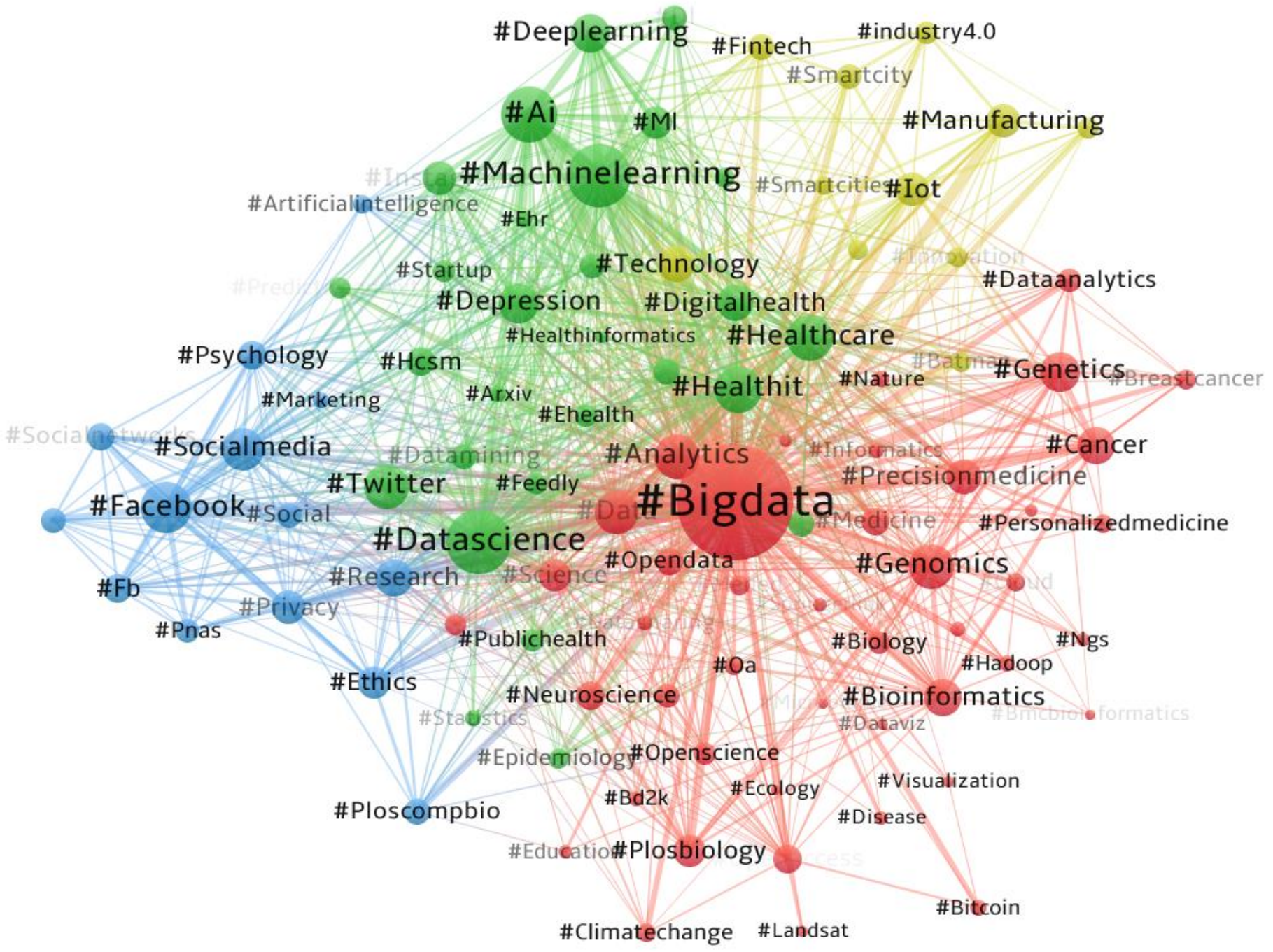

**Fig. 5** Cluster map of high-frequency hashtags.

Table 4 lists the top 10 high-frequency hashtags and their occurrences. "#Bigdata" tops the list with over 4000 (9.65%) tweets, contributing to almost 10% of all the information provided by hashtags, far ahead of the others. Following is "#Datascience" with frequency around 500, which is also a popular concept in recent years. It primarily involves the processes for extracting and discerning valuable knowledge from complex data, as well as the development and use of related tools (Leek 2013; Waller and Fawcett 2013), so is quite associated with “Big Data”. The third and fourth topics are both technical terms of emerging and popular technologies for data mining and data analysis (“#MachineLearning” and “#AI”). Moreover, as mentioned above, health care relevant topics (“#Health”, “#Genomics”, and “#Healthcare”) are also prominent among Twitter users. In addition, compared with top-10 terms, the coverage of top-10 hashtags in tweets is relatively low, indicating a broader range of topics discussed by the Twitter audiences around Big Data publications.

**Table 4** Top-10 high-frequency hashtags

| | Hashtags | Occurrences | Share (%) | Share of tweets (%) | Cluster |
|---|---|---|---|---|---|
| 1 | #Bigdata | 4,088 | 9.87 | 9.65 | 1 |
| 2 | #Datascience | 498 | 0.96 | 1.18 | 2 |
| 3 | #MachineLearning | 419 | 0.76 | 0.99 | 2 |
| 4 | #AI | 253 | 0.68 | 0.60 | 2 |
| 5 | #Analytics | 238 | 0.66 | 0.56 | 1 |
| 6 | #Facebook | 237 | 0.57 | 0.56 | 3 |
| 7 | #Data | 228 | 0.55 | 0.54 | 1 |
| 8 | #Health | 205 | 0.50 | 0.48 | 2 |
| 9 | #Genomics | 203 | 0.49 | 0.48 | 1 |
| 10 | #Healthcare | 195 | 0.47 | 0.46 | 2 |

## Similarity measurement

After simple integration, for example, unifying the plural and singular forms of words, replacing abbreviations with full names, removing hyphens, etc., the author keywords, textual terms, and hashtags appeared in the Figs 2-5 can form a list of 235 distinct topics. In other words, the topic list covers all the top-100 author keywords, terms, and hashtags, ranging in frequency from one to three (with one meaning that the given topic only appears in one group, while three implies that it occurs in all the three groups as a common topic). All the 235 topics and their occurrence in each group can be seen in Table 8 in the *Appendix*.

Venn diagram in Fig. 6 shows the layout of the 235 topics divided into seven parts with different colors. The numbers of topics in each part have been marked in the figure. Taking the group of author keywords (red) as an example, the 100 author keywords are separated into four parts: 60 occur as keywords only, 11 are in common with both other two groups (i.e. hashtags, blue, and terms, green), 27 also appear in hashtags and two in terms. Table 5 provides the result of the similarity measurement between group pairs. Of all the topics, only 11 (5.15%) are duplicated in all the three groups, demonstrating that nearly one in ten of the academic topics from Big Data publications are also highly concerned by the online audiences. Hashtags and author keywords have the largest number of common topics and the largest cosine similarity (38, 0.38). Following are hashtags and terms (25, 0.25), whereas terms and author keywords have the least similarity (13, 0.13).

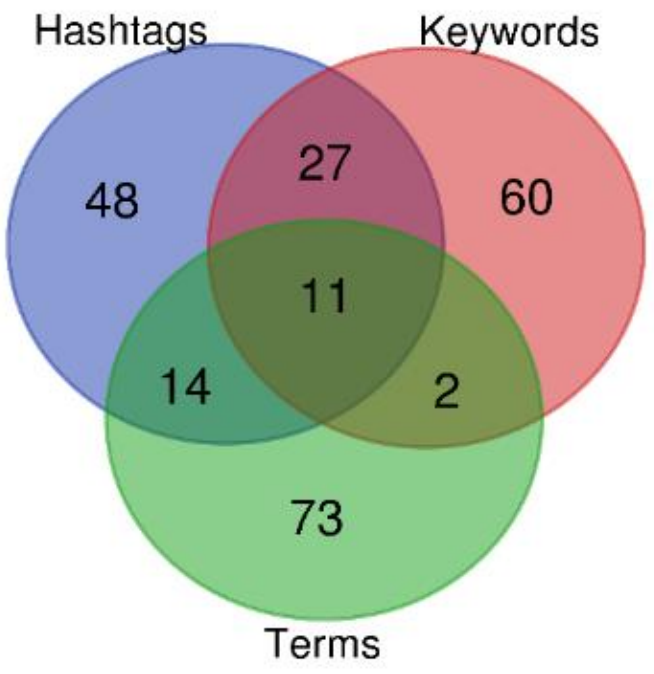

**Fig. 6** Venn diagram of topic sets.

**Table 5** Cosine similarity of topic pairs

| | Terms | Keywords | Hashtags |
|---|---|---|---|
| **Terms** | | 0.1300 | 0.2500 |
| **Keywords** | | | 0.3800 |
| **Hashtags** | | | |

By breaking down the second audience group into four sub-groups according to the platforms they used (Blogs, News, Policy, and Wikipedia), we further investigated the topic similarity among them. The results are shown in Table 6 and Fig. 7. Blogs and News have the strongest similarity (0.9587) due to their larger number of topics included, which increases the possibility of having a common topic. Overall, News covers all the terms in Blogs and Wikipedia, and almost all the terms in Policy (27/28). The similarity between Blogs and Wikipedia ranks second (0.7703), and all the terms in Wikipedia are covered by those in Blog. Policy and Wikipedia are the least similar (0.4629) on topics among these platforms, which means they have different semantic orientations in the terms they used. Besides, when considering all the six groups together, topic sets from Blogs and News also have a higher degree of similarity to those from Twitter and publications (see Fig. 14 and Table 9 in *Appendix*).

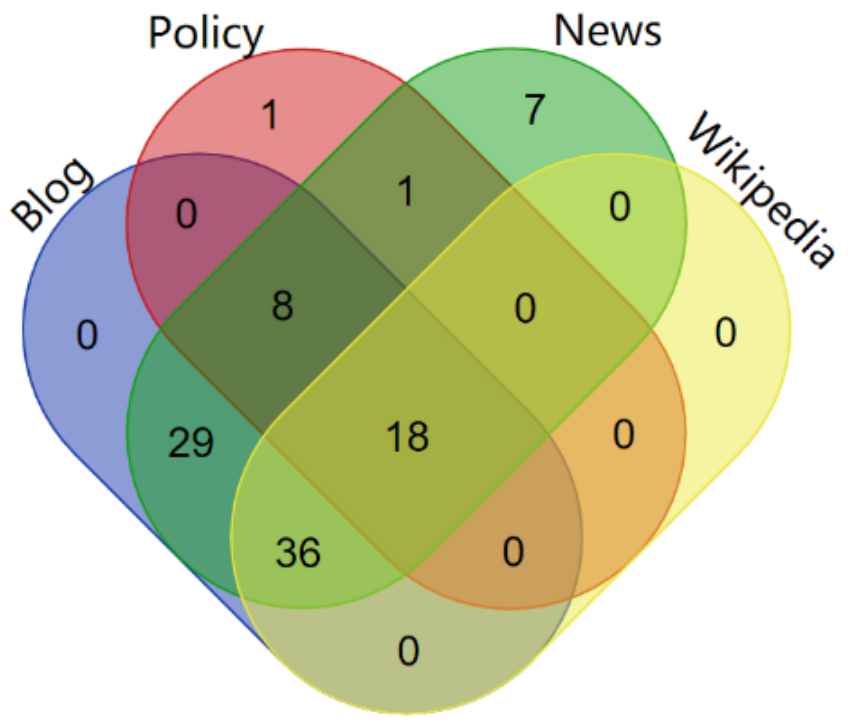

**Fig. 7** Venn diagram of topic sets.

**Table 6** Cosine similarity of topic pairs

| | Blog (91) | Policy (28) | News (99) | Wikipedia (54) |
|---|---|---|---|---|
| **Blog (91)** | | 0.5151 (26) | 0.9587 (91) | 0.7703 (54) |
| **Policy (28)** | | | 0.5128 (27) | 0.4629 (18) |
| **News (99)** | | | | 0.7385 (54) |
| **Wikipedia (54)** | | | | |

## Comparison of topic sets

*Common and different topics*

The word cloud[9] in Fig. 8 displays the 11 common topics (KTH) of the three groups, that is, the central part in Fig 6. The size of each word (topic) is based on its total frequency of occurrence in the three groups. Therefore, the bigger the size, the more frequently it appears, and the more attention it has received from both academic authors and online audiences. Apparently, the 11 common topics illustrate that emerging technologies, especially “Artificial intelligence” and “Machine learning”, are highly relevant terms in Big Data publications and online discussions as well, which are quite conspicuous in this figure. In fact, as new technologies that require a considerable volume of information in the form of big data to function, practical applications of Artificial Intelligence (AI) and Machine Learning (ML) have been on the rise in all business areas and daily life (Zhang et al. 2019). Therefore, they are common topics both in academia and online communities. Besides, some general topics which are closely related to the development of human society (e.g., “Health care”, “Climate change” and “Privacy”) also have been frequently used, highlighting the opportunities and challenges we are facing in the era of Big Data.

[9] The online platform WordItOut (https://worditout.com/word-cloud/create) is used for showing the word cloud layouts in our study.

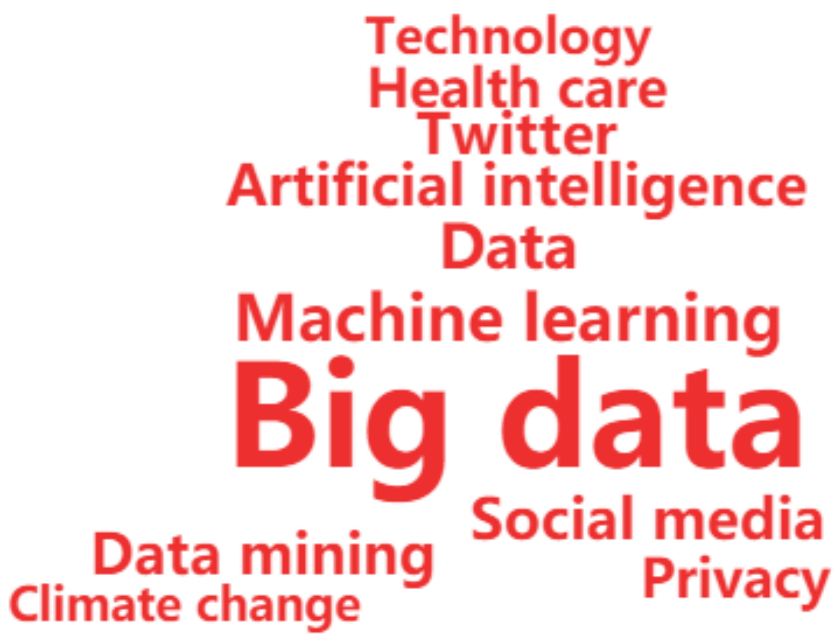

**Fig. 8** Common topics of scholars and audiences (KTH, 11).

Further observation on the rankings (i.e. importance) of the common topics in each group reveals different degrees of attention of these topics by the subjects (Table 7). The numbers in the table represent the order of each topic in different ranking groups. Taking the ranking of author keywords as the baseline, the arrows represent the change trend of rankings of these topics in other two groups. Compared with the baseline, 6 topics (i.e. “Data”, “Artificial intelligence”, “Twitter”, “Health care”, “Technology” and “Climate change”) have increased their status significantly in hashtag ranking on Twitter, among which “Data” and “Artificial intelligence” jump from the middle in keywords to top-10 in hashtags, while “Climate change” is the biggest mover in the list (from 95$^{th}$ to 49$^{th}$). Three topics (i.e. “Data”, “Artificial intelligence”, and “Climate change”) have also improved their positions in term ranking. In addition, more topics (8) have slipped places to varying degrees in the ranking of terms than in hashtags (3), among which the high frequency of “Social media”, “Data mining” and “Privacy” as author keywords decreased in their ranking in the online discussions. Besides, “Big Data” and “Machine learning” keep ahead in ranking of hashtags with wide mention, but not the case in the other platforms in general.

**Table 7** Rankings of common topics in the three groups

| | Common Topics | Ranking in | | | | |
|---|---|---|---|---|---|---|
| | | Keywords | Hashtags | | Terms | |
| 1 | Big data | 1 | 1 | — | 2 | ↓ |
| 2 | Machine learning | 3 | 3 | — | 63 | ↓ |
| 3 | Data mining | 4 | 32 | ↓ | 90 | ↓ |
| 4 | Social media | 6 | 14 | ↓ | 59 | ↓ |
| 5 | Privacy | 10 | 17 | ↓ | 44 | ↓ |

| | Common Topics | Ranking in | | | | |
|---|---|---|---|---|---|---|
| | | Keywords | Hashtags | | Terms | |
| 6 | Twitter | 13 | 11 | ↑ | 37 | ↓ |
| 7 | Health care | 31 | 10 | ↑ | 79 | ↓ |
| 8 | Artificial intelligence | 32 | 4 | ↑ | 75 | ↓ |
| 9 | Data | 45 | 7 | ↑ | 3 | ↑ |
| 10 | Technology | 75 | 42 | ↑ | 26 | ↑ |
| 11 | Climate change | 95 | 49 | ↑ | 71 | ↑ |

As for the different topics of publications or audiences (i.e. K or H/T/HT), the *pure academic focus (K)* are more technical and professional, of which most are scientific jargon not easily understood by the public or ordinary laymen, such as “Hdfs” (the Hadoop Distributed File System), “Surveillance” or “GPU” (the Graphics Processing Unit). Other business-related topics have also been the focus of authors but not online audiences, for instance, “Business intelligence”, “Resource allocation” and “Supply chain management”, which may to some extent indicate the prosperity of information economy with the development of Big Data applications and the Internet of Things (Fig. 9).

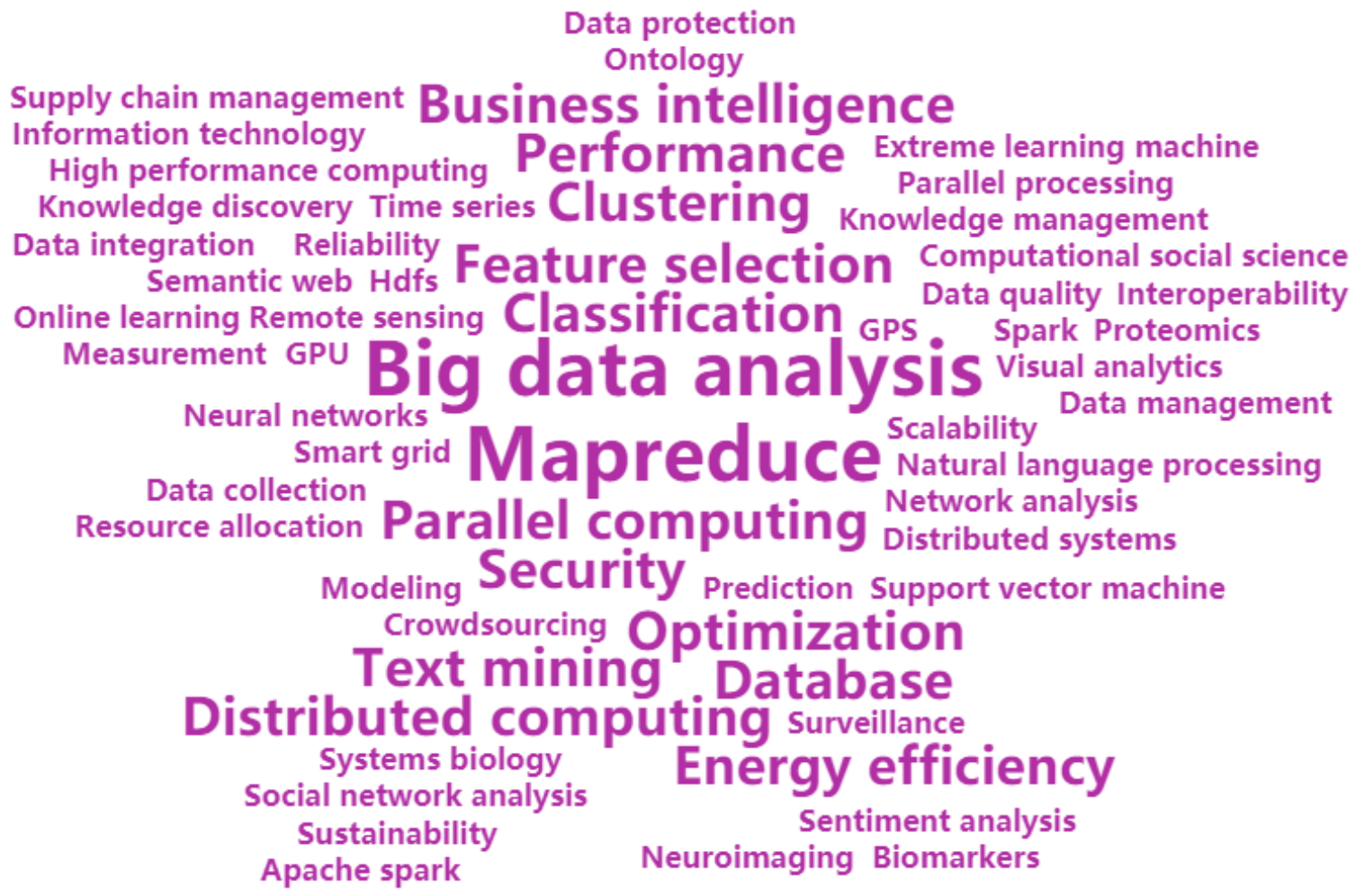

**Fig. 9** Pure academic topics of publications (K, 60).

With regards to the *pure audience topics* (H/T/TH), we further divide them into three parts based on their occurrences in the two audiences groups: *pure hashtags* (H, 48, 35.1%, orange), *pure terms* (T, 73, 54.5%, blue), and the *common ones* (HT, 14, 10.4%, green). Comparison of pure hashtags and terms provides evidence that Twitter audiences discuss more topics related to academic research in various disciplines, such as “Neuro-science”, “Genetics”, “Plosbiology” and “Gahitec”, of which biology and health are the most widely covered themes. As mass media disseminating social hotspots and news anecdotes, Blogs, Policy, Wikipedia, and News tend to report general social events or technological advances, so the users’ concerns are generally less technical and more comprehensible (e.g., “Study”, “Researcher” and “Scientist”). Additionally, the common topics between these two audiences groups emphasize that, in addition to scientific research as an essential information source, mental health-related event draws great attention in the online community at present (Fig. 10).

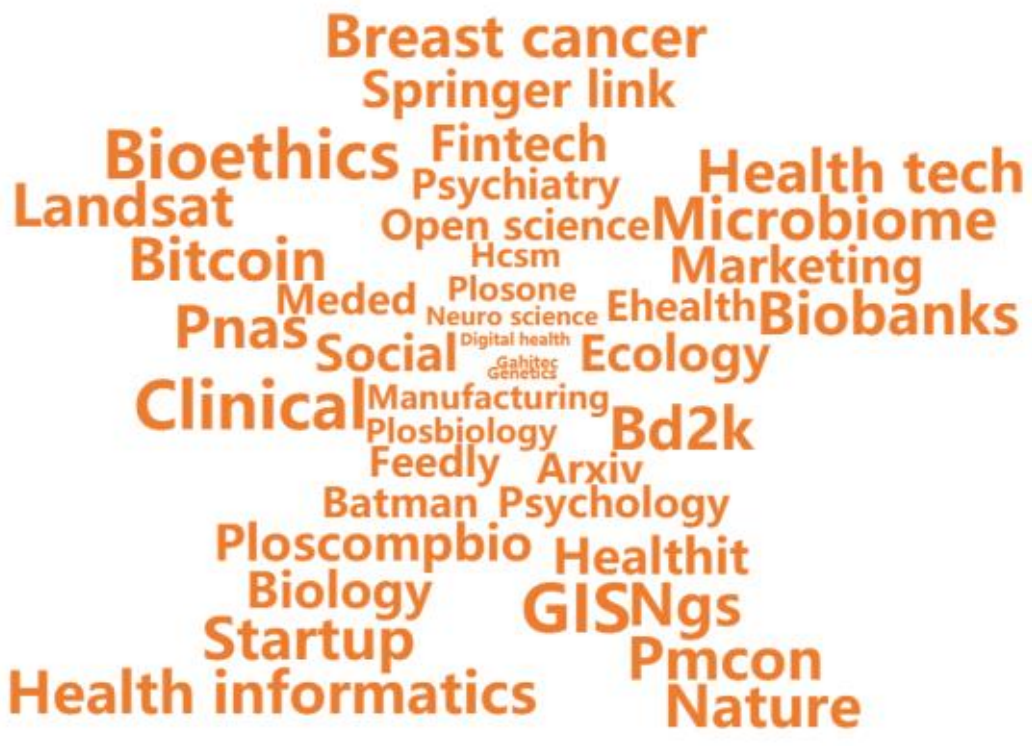

**Pure hashtags (H, 48)**

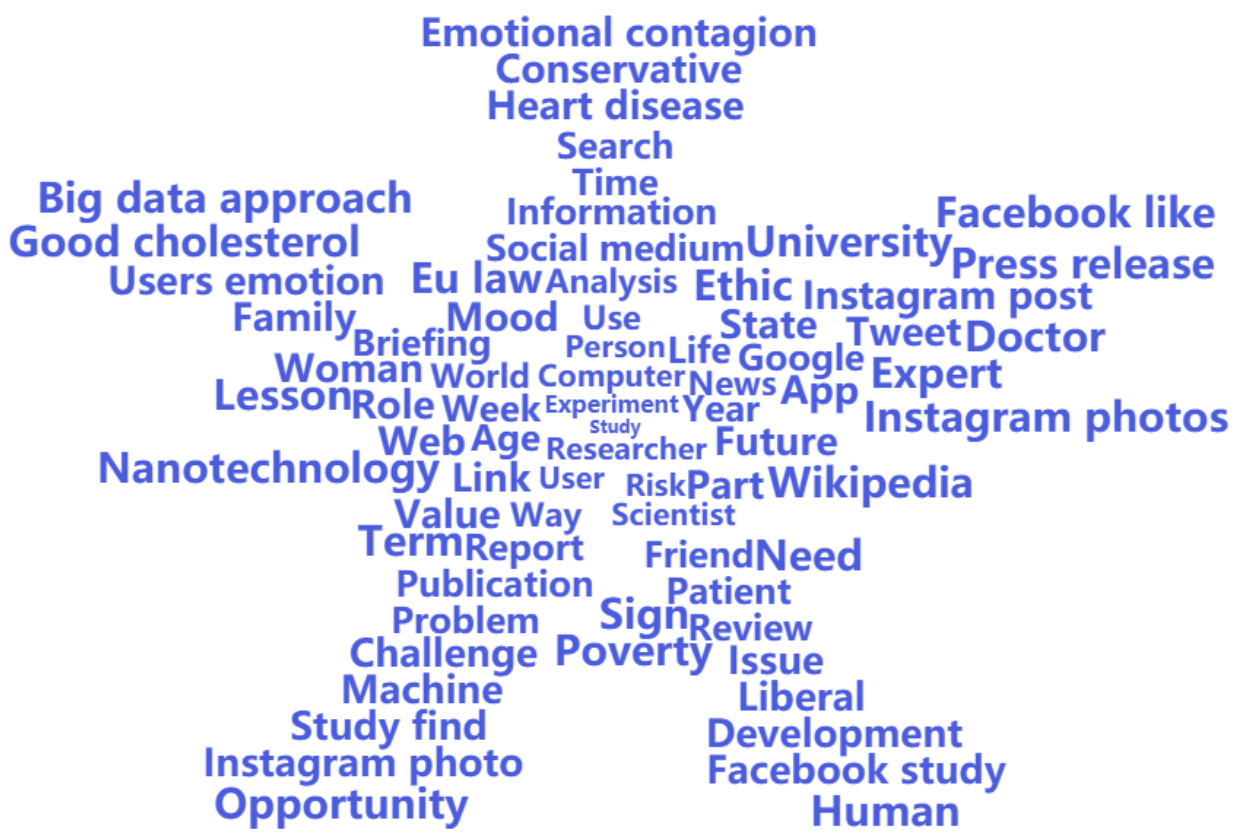

**Pure terms (T, 73)**

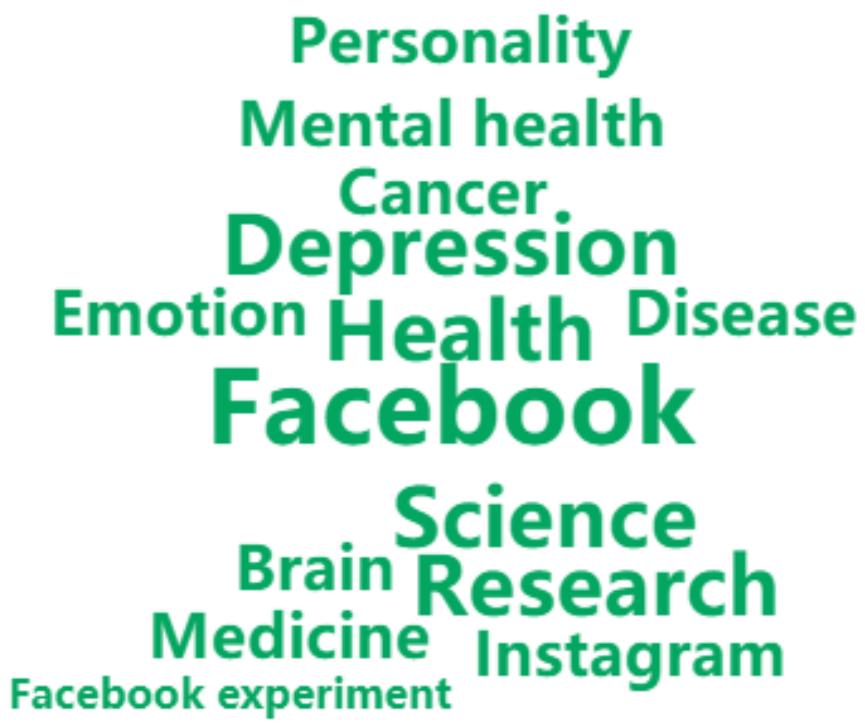

**Common topics (HT, 14)**

**Fig. 10** Pure online audience topics (135) divided into three parts.

*Shift of academic topics*

The relationship between the online posts and the mentioned Big Data publications enables us to establish two-way linkages between author keywords and audience terms (i.e. hashtags and terms). In this section, we examined the top-5 highly-mentioned author keywords and their linked audience topics on Twitter and the other four platforms, respectively. Such one-to-many linkages can reflect not only the diverse discussions but also the shift pattern around the specific topics among social media users from a thematic perspective.

Fig. 11 shows the top-5 highly-mentioned author keywords (green) by audiences on Blogs, News, Policy and Wikipedia, as well as the top-5 text terms (red, signaled with "T:") with most links to the keywords. The size of topics and the thickness of lines are both based on the frequency of occurrence. In other words, the bigger the size of the nodes, the thicker the lines connected to it, the higher the frequency of the topic. Obviously, "T: Facebook" is the closest audience concern to these academic topics, which can be mirrored by its high link rate with the academic topics (4/5). Technically, "T: Facebook" contributes nearly 14% of the mention rate to "Social media", and approximately 8% to "Data mining" and also "Big data". Moreover, "Machine Learning" and "Social media" are more often used to discuss topics related to mental health by the audiences (e.g., "T: Mental health" and "T: Depression"), while "Privacy" has been interpreted more concretely (e.g., "T: Preserving privacy" and "T: Medical privacy").

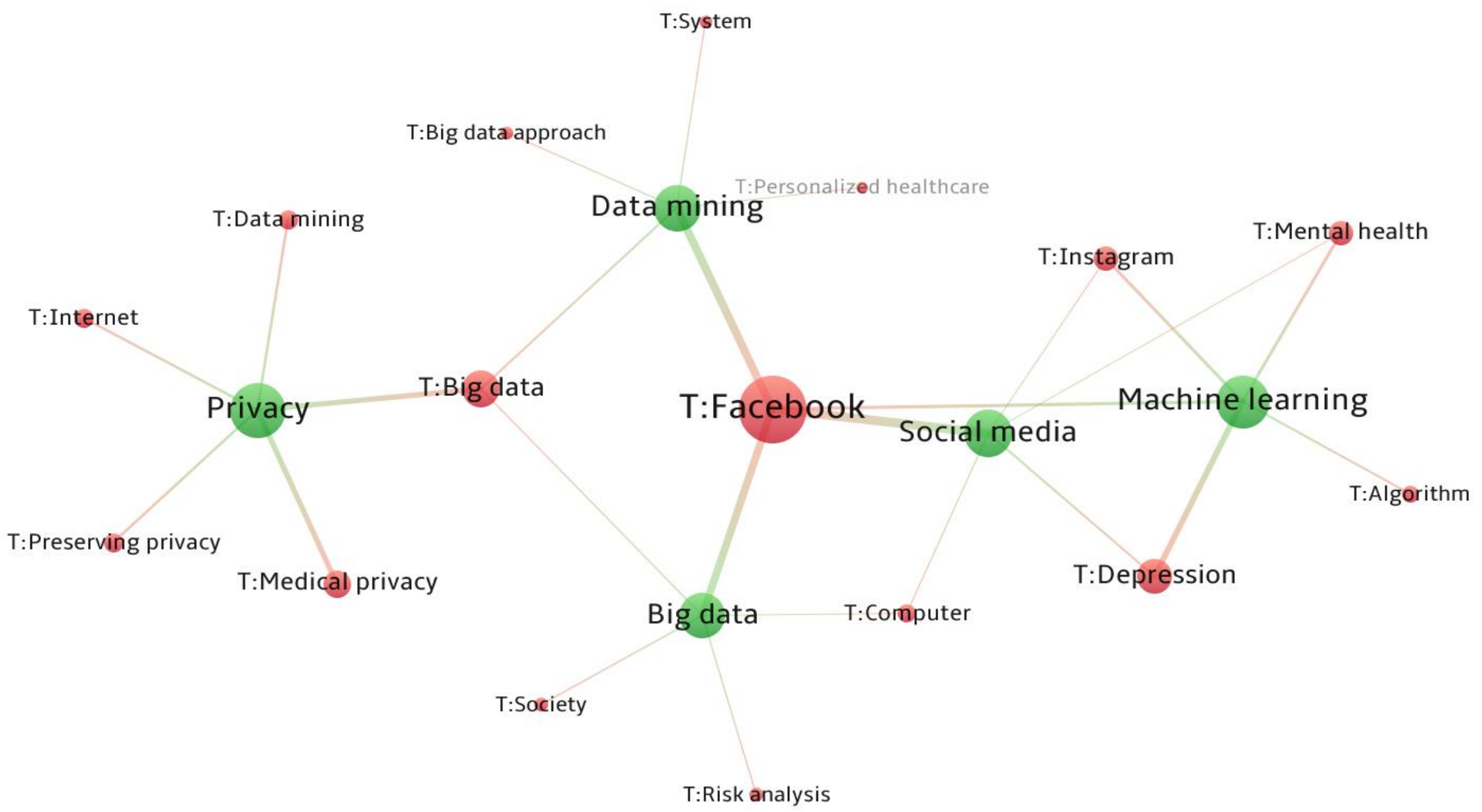

**Fig. 11** Top-5 highly mentioned author keywords and top-5 terms with most links to them.

The same approach is also applied for the top-5 highly-mentioned author keywords in tweets (green) and their linked hashtags (red, signaled with #). The result is displayed in Fig. 12. Compared with the network in Fig. 11, this network has a better connectivity with more items connected to each other. Moreover, “#Bigdata” replaces the central position of “T: Facebook” in terms, linked to all the five academic topics. The frequently mentioned author keyword “Big data” is connected mostly with “#Bigdata” and “#Datascience” on Twitter. The relationship between these two concepts is also a popular debate among scholars in various fields (e.g., Kacfah et al. 2015; Park and Leydesdorff 2013; Phillips 2017; Gupta and Rani 2018), and this analysis shows that these two concepts are also popular among Twitter users. Besides the application of data analysis methods in the field of biomedicine, with more appeals about open data and data sharing, “#Privacy” is also a significant concern closely related to “Big Data”. As technical terms, “Data mining” and “Machine learning” are usually connected with technologies via hashtags, for instance, “#Machinelearning”, “#ArtificialIntelligence”, and “#Deeplearning”, suggesting that Twitter audiences are also quite concerned about the development of core technologies. Discussions related to social media and social networks focus on specific platforms like “#Facebook” and “#Twitter”, as well as general issues, such as “#Healthcare” and “#Privacy”.

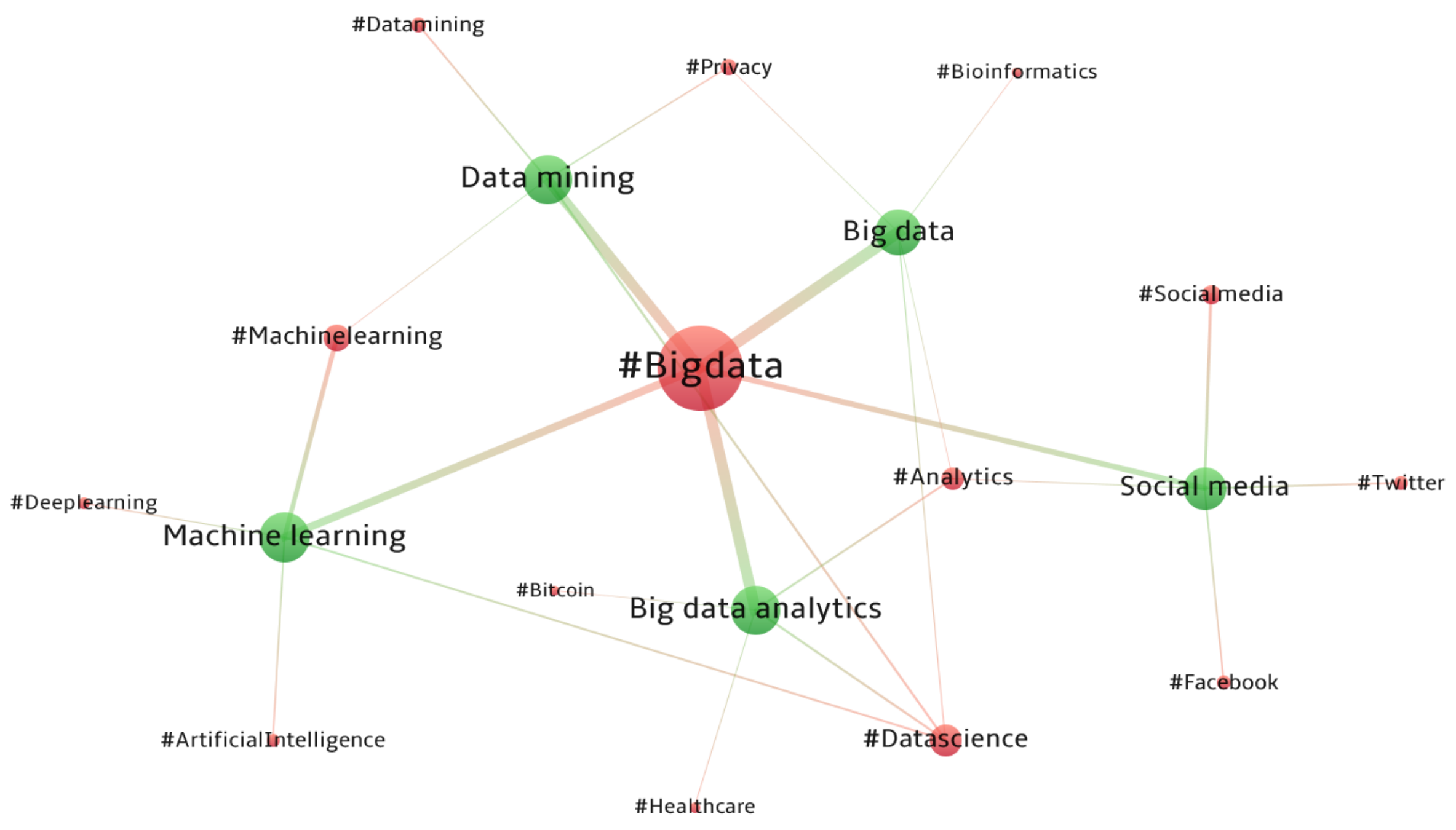

**Fig. 12** Top-5 highly mentioned author keywords and top-5 hashtags with most links to them.

## Discussion and Conclusion

Unlike most previous research on SMM focusing on the impact of publications on social media and their correlation with citation or mention counts, in this paper, we study how academic topics in the research area of Big Data have been transformed across different altmetric sources. More specifically, we examined and measured the degree of similarity between the sets of terms used by publication authors, and the terms used by their online audiences across different platforms. We argue that this approach can open up a new research window to study the role of online audiences in the dissemination of academic topics from academia to the online community from a more semantic perspective.

Based on high-frequency author keywords from publications and textual terms from online events, the main topics in Big Data publications across different communities have been identified separately. It is revealing that there exist different thematic tendencies among these groups. Big Data authors pay more attention to technology development than their online audiences. This is shown by a large cluster of technical terms among the author keywords, like "Cloud computing", "Mapreduce", and "Machine learning". This technical orientation can also be observed among Twitter audiences. Terms used in blogs and news show an interest in popularizing scientific research and discovery, as well as in interpersonal relations. Policy documents tend to focus on more general and political issues, while those on Wikipedia are more related to the application of data analysis technologies on the Internet. Besides, core technologies (i.e. "Artificial Intelligence" and "Machine Learning") and some general issues (i.e. "health care", "climate change" and "social media") are the most important common topics among both authors and online audiences.

Similarity metrics provide us with a more numeric description of the degree of differences in user interests across different platforms, showing that Twitter audiences and Big Data authors have more common topics of interest than the other audience groups. Several possible explanations for this stronger similarity between Twitter hashtags and author keywords are taken into account. First, the substantial number of mechanical interactions with publications on Twitter makes it easy to generate tweets by clicking on the Twitter icon on the pages of journal articles, thus greatly increases the original content from these papers in the online discussion among Twitter users (Robinson-Garcia et al. 2017). Secondly, the large amount of retweets produced by simply copying the original tweets (boyd et al. 2010) increase the repetition rate of hashtags used on Twitter. Besides, there is a large group of scholars with publications included in the WoS database who are also active on Twitter (Costas et al. 2017; Yu et al. 2019), which means that these scholars may use the same academic terms in their Twitter use of hashtags.

When it comes to the other audience group, Blog and News users have the largest degree of similarity in the terms they used to introduce and interpret Big Data publications, while Policy and Wikipedia show the lowest. One reason that cannot be ignored is that science journalists are a large group of actors in science blogging, aiming at explaining science broadly and educate readers (Bartling and Friesike 2014), so they may post the same or similar content in blogs and news (Fraumann et al. 2015). In our dataset, 97 events from Blog and News have the same headlines, which improves the degree of similarity between the two topic sets, while there are almost no identical titles between other platforms.

Further investigation into the pure academic focus offer an insight of the lower adoption of the more technical and professional terminologies by the online audiences, probably because these more technical terminology are not easily understood by the public and the non-specialists. On the other hand, the pure hashtags and terms that are not commonly used by the authors, can be regarded as a form of expansion and reinterpretation of the academic topics around the research area of Big Data by social media audiences. More specifically, Twitter users have turned to discussing or linking to topics involving medicine, biology, humanities, social sciences and other disciplines, demonstrating to some extent the widespread distribution of its users and the diversity of their opinions and views. Blogs, policy documents, Wikipedia articles, and news tend to report more general topics with terms that are less professional and easy to understand, somehow introducing a more people's daily life perspective.

In conclusion, our case study has proven that there are indeed (dis)similarities between the topics highlighted by authors in their papers and how they are discussed by online audiences. Overall, it can be concluded that the online users tend to mention topics that are more social and general. Simultaneously, they can help to further interpret, spread, and diversify academic topics, contributing to relate the scientific research with more practical problems.

*Limitations of this study*

The research presented in this study is also bound by some limitations that deserve further discussion. First, we only study papers indexed in WoS with limited types of article, review and letter, which means a large volume of proceeding papers and other papers not included in WoS are excluded. Besides, since a small part (about 10%) of the papers in our dataset do not have a DOI, more comparable identifiers, like arXiv ID or PMID, should be adopted for matching papers with those mentioned on the altmetric sources. Third, considering the difficulty of data analysis and processing, only English events (the overwhelming majority) are taken into account in this paper. In addition, since the Wikipedia titles are just the name of the entry, we chose the first sentence of the entry for term extraction, under the assumption that this sentence tends to provide a preliminary definition of the entry. However, there may be also conceptual differences between the first sentence of summary and the titles of blogs, news and policy documents that need to be studied in future research. Regarding Twitter, we only focused on comparing hashtags and author keywords. This choice has the advantage that we are comparing conceptually special features in both articles and tweets (i.e. hashtags are intendedly “selected” keywords by the Twitter users in order to frame the tweet, conceptually similar to the author keywords of publications). In future research, it would be relevant to also study the full text of tweets in order to better characterize the type of engagement of tweeters with the contents of the publications.

Finally, we would like to point out that there is a wide variability in the use and uptake of social media tools across different communities. Much of the published research has sought to identify factors of differentiation, such as age, academic level, gender, discipline, country and language, as well as the technical level of scholars using such tools (e.g., Nicholas et al. 2014; Mansour 2015; Larivière et al. 2013; Priem et al. 2012; Cronin and Sugimoto 2015). Therefore, according to these factors put forward in previous research, follow-up studies can be conducted to further analyze the (dis)similarity in the degree of attention and promotion of academic topics among different user groups in the online communities.

## Acknowledgements

Xiaozan Lyu was supported by China Scholarship Council (CSC Student ID 201806320214), and the National Science Foundation of China (NSFC, Grant Number: 71843012). Rodrigo Costas was partially supported by funding from the DST-NRF Centre of Excellence in Scientometrics and Science, Technology and Innovation Policy (SciSTIP) (South Africa). The authors thank an anonymous reviewer for the insightful comments of an early version of the paper. The authors also acknowledge Altmetric.com for providing the altmetric data of scientific publications for this study.

# Appendix

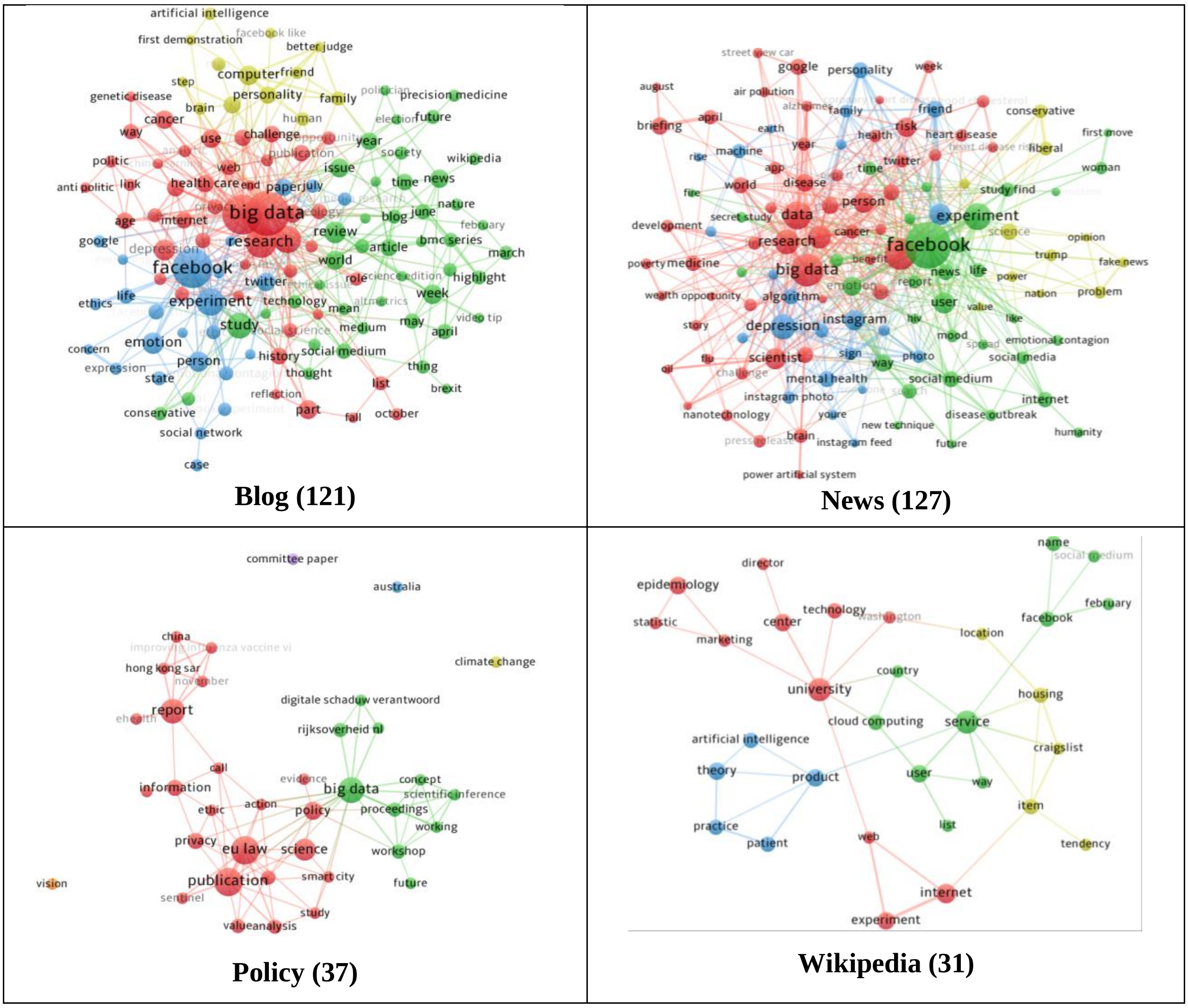

**Fig. 13** Term maps of Blog, News, Policy and Wikipedia. The minimum number of occurrence for being plotted is 3 for terms from blogs and news, and 2 for terms from Policy documents and Wikipedia articles.

**Table 8** 235 topics appeared in at least one group and their occurrence in each group.

| | Topics | Type | Term (T) | | | | | Keywords (K) | Hashtags (H) | Total |
|---|---|---|---|---|---|---|---|---|---|---|
| | | | Total | Blog | News | Policy | Wiki | | | |
| 1 | Big data | KTH | 186 | 75 | 97 | 10 | 4 | 2960 | 4088 | 7234 |
| 2 | Machine learning | KTH | 16 | 3 | 11 | | 2 | 264 | 315 | 595 |
| 3 | Facebook | TH | 276 | 59 | 214 | | 3 | | 280 | 556 |
| 4 | Data science | KH | | | | | | 63 | 397 | 460 |
| 5 | Data | KTH | 155 | 60 | 69 | 8 | 18 | 37 | 228 | 420 |
| 6 | Cloud computing | KH | | | | | | 328 | 21 | 349 |
| 7 | Social media | KTH | 17 | 6 | 11 | | | 163 | 168 | 348 |
| 8 | Artificial intelligence | KTH | 13 | 4 | 7 | | 2 | 44 | 274 | 331 |
| 9 | Internet of things | KH | | | | | | 198 | 124 | 322 |
| 10 | Data mining | KTH | 11 | 1 | 4 | | 6 | 244 | 62 | 317 |

| | Topics | Type | Term (T) | | | | | Keywords (K) | Hashtags (H) | Total |
|---|---|---|---|---|---|---|---|---|---|---|
| | | | **Total** | Blog | News | Policy | Wiki | | | |
| 11 | Analytics | KH | | | | | | 79 | 238 | 317 |
| 12 | Twitter | KTH | 23 | 10 | 13 | | | 78 | 189 | 290 |
| 13 | Privacy | KTH | 21 | 7 | 8 | 4 | 2 | 107 | 133 | 261 |
| 14 | Genomics | KH | | | | | | 54 | 203 | 257 |
| 15 | Research | TH | 112 | 46 | 52 | 5 | 9 | | 140 | 252 |
| 16 | Health care | KTH | 12 | 9 | 3 | | | 44 | 195 | 251 |
| 17 | Precision medicine | KH | | | | | | 65 | 181 | 246 |
| 18 | Bioinformatics | KH | | | | | | 66 | 169 | 235 |
| 19 | Mapreduce | K | | | | | | 232 | | 232 |
| 20 | Big data analysis | K | | | | | | 227 | | 227 |
| 21 | Health | TH | 21 | 7 | 11 | 2 | 1 | | 205 | 226 |
| 22 | Deep learning | KH | | | | | | 70 | 135 | 205 |
| 23 | Open access | H | | | | | | | 189 | 189 |
| 24 | Science | TH | 68 | 33 | 18 | 7 | 10 | | 111 | 179 |
| 25 | Hadoop | KH | | | | | | 139 | 35 | 174 |
| 26 | Depression | TH | 66 | 13 | 53 | | | | 108 | 174 |
| 27 | Data analytics | KH | | | | | | 132 | 31 | 163 |
| 28 | Smart city | KH | | | | | | 78 | 75 | 153 |
| 29 | Open data | KH | | | | | | 48 | 100 | 148 |
| 30 | Cancer | TH | 26 | 10 | 15 | | 1 | | 120 | 146 |
| 31 | Technology | KTH | 29 | 5 | 20 | 1 | 3 | 25 | 81 | 135 |
| 32 | Ethics | KH | | | | | | 53 | 80 | 133 |
| 33 | Study | T | 131 | 22 | 99 | 2 | 8 | | | 131 |
| 34 | Mental health | TH | 19 | | 19 | | | | 99 | 118 |
| 35 | Gahitec | H | | | | | | | 117 | 117 |
| 36 | Genetics | H | | | | | | | 100 | 100 |
| 37 | Digital health | H | | | | | | | 98 | 98 |
| 38 | Instagram | TH | 38 | 4 | 33 | | 1 | | 59 | 97 |
| 39 | Epidemiology | KH | | | | | | 49 | 42 | 91 |
| 40 | Medicine | TH | 31 | 11 | 18 | 1 | 1 | | 59 | 90 |
| 41 | Algorithm | TK | 22 | 3 | 18 | | 1 | 64 | | 86 |
| 42 | Experiment | T | 86 | 26 | 58 | | 2 | | | 86 |
| 43 | Climate change | KTH | 14 | 1 | 11 | 2 | | 22 | 42 | 78 |
| 44 | Visualization | KH | | | | | | 56 | 20 | 76 |
| 45 | Personalized medicine | KH | | | | | | 37 | 39 | 76 |
| 46 | Social networks | KH | | | | | | 42 | 33 | 75 |
| 47 | Neuro science | H | | | | | | | 74 | 74 |
| 48 | Emotion | TH | 53 | 18 | 34 | | 1 | | 20 | 73 |
| 49 | Electronic health records | KH | | | | | | 36 | 31 | 67 |
| 50 | Internet | TK | 33 | 8 | 19 | 2 | 4 | 33 | | 66 |
| 51 | Predictive analytics | KH | | | | | | 41 | 24 | 65 |
| 52 | Manufacturing | H | | | | | | | 65 | 65 |
| 53 | Plosone | H | | | | | | | 63 | 63 |

| | Topics | Type | Term (T) | | | | | Keywords (K) | Hashtags (H) | Total |
|---|---|---|---|---|---|---|---|---|---|---|
| | | | **Total** | Blog | News | Policy | Wiki | | | |
| 54 | Clustering | K | | | | | | 62 | | 62 |
| 55 | Data sharing | KH | | | | | | 39 | 23 | 62 |
| 56 | Public health | KH | | | | | | 21 | 40 | 61 |
| 57 | Plosbiology | H | | | | | | | 61 | 61 |
| 58 | HCSM | H | | | | | | | 61 | 61 |
| 59 | Classification | K | | | | | | 60 | | 60 |
| 60 | Cloud | KH | | | | | | 30 | 30 | 60 |
| 61 | Statistics | KH | | | | | | 28 | 32 | 60 |
| 62 | Innovation | KH | | | | | | 29 | 29 | 58 |
| 63 | Disease | TH | 35 | 7 | 24 | | 4 | | 23 | 58 |
| 64 | Brain | TH | 20 | 4 | 16 | | | | 36 | 56 |
| 65 | Informatics | KH | | | | | | 31 | 23 | 54 |
| 66 | Person | T | 54 | 13 | 34 | | 7 | | | 54 |
| 67 | Feedly | H | | | | | | | 53 | 53 |
| 68 | Psychology | H | | | | | | | 53 | 53 |
| 69 | Open science | H | | | | | | | 53 | 53 |
| 70 | Researcher | T | 53 | 9 | 43 | | 1 | | | 53 |
| 71 | Text mining | K | | | | | | 51 | | 51 |
| 72 | Arxiv | H | | | | | | | 51 | 51 |
| 73 | Computer | T | 50 | 12 | 36 | | 2 | | | 50 |
| 74 | Security | K | | | | | | 49 | | 49 |
| 75 | Feature selection | K | | | | | | 48 | | 48 |
| 76 | Scientist | T | 48 | 9 | 37 | | 2 | | | 48 |
| 77 | Meded | H | | | | | | | 48 | 48 |
| 78 | Psychiatry | H | | | | | | | 48 | 48 |
| 79 | Personality | TH | 28 | 9 | 19 | | | | 19 | 47 |
| 80 | User | T | 46 | 6 | 37 | | 3 | | | 46 |
| 81 | Risk | T | 46 | 6 | 38 | 2 | | | | 46 |
| 82 | Optimization | K | | | | | | 45 | | 45 |
| 83 | Education | KH | | | | | | 24 | 21 | 45 |
| 84 | Batman | H | | | | | | | 45 | 45 |
| 85 | E health | H | | | | | | | 44 | 44 |
| 86 | Industry 4.0 | KH | | | | | | 22 | 21 | 43 |
| 87 | Energy efficiency | K | | | | | | 42 | | 42 |
| 88 | Parallel computing | K | | | | | | 41 | | 41 |
| 89 | Database | K | | | | | | 41 | | 41 |
| 90 | Performance | K | | | | | | 40 | | 40 |
| 91 | Distributed computing | K | | | | | | 40 | | 40 |
| 92 | Business intelligence | K | | | | | | 40 | | 40 |
| 93 | Ploscompbio | H | | | | | | | 40 | 40 |
| 94 | Fintech | H | | | | | | | 40 | 40 |
| 95 | Prediction | K | | | | | | 39 | | 39 |
| 96 | Crowdsourcing | K | | | | | | 39 | | 39 |
| 97 | Analysis | T | 38 | 10 | 21 | 3 | 4 | | | 38 |

| | Topics | Type | Term (T) | | | | | Keywords (K) | Hashtags (H) | Total |
|---|---|---|---|---|---|---|---|---|---|---|
| | | | **Total** | Blog | News | Policy | Wiki | | | |
| 98 | Springer link | H | | | | | | | 37 | 37 |
| 99 | News | T | 37 | 10 | 26 | 1 | | | | 37 |
| 100 | Facebook experiment | TH | 11 | 5 | 6 | | | | 26 | 37 |
| 101 | Remote sensing | K | | | | | | 36 | | 36 |
| 102 | Social | H | | | | | | | 36 | 36 |
| 103 | Health IT | H | | | | | | | 36 | 36 |
| 104 | Systems biology | K | | | | | | 35 | | 35 |
| 105 | Ecology | H | | | | | | | 35 | 35 |
| 106 | Surveillance | K | | | | | | 34 | | 34 |
| 107 | Spark | K | | | | | | 34 | | 34 |
| 108 | Ontology | K | | | | | | 34 | | 34 |
| 109 | Data protection | K | | | | | | 34 | | 34 |
| 110 | Use | T | 34 | 10 | 18 | 3 | 3 | | | 34 |
| 111 | Social network analysis | K | | | | | | 32 | | 32 |
| 112 | Sentiment analysis | K | | | | | | 32 | | 32 |
| 113 | Biology | H | | | | | | | 32 | 32 |
| 114 | Way | T | 32 | 7 | 23 | | 2 | | | 32 |
| 115 | Sustainability | K | | | | | | 31 | | 31 |
| 116 | World | T | 31 | 11 | 18 | | 2 | | | 31 |
| 117 | Marketing | H | | | | | | | 31 | 31 |
| 118 | Scalability | K | | | | | | 30 | | 30 |
| 119 | Data quality | K | | | | | | 30 | | 30 |
| 120 | Social medium | T | 30 | 7 | 20 | 1 | 2 | | | 30 |
| 121 | Semantic web | K | | | | | | 29 | | 29 |
| 122 | Biomarkers | K | | | | | | 29 | | 29 |
| 123 | NGS | H | | | | | | | 29 | 29 |
| 124 | GPS | K | | | | | | 29 | | 29 |
| 125 | Time series | K | | | | | | 28 | | 28 |
| 126 | Apache spark | K | | | | | | 28 | | 28 |
| 127 | Computational social science | K | | | | | | 28 | | 28 |
| 128 | Visual analytics | K | | | | | | 27 | | 27 |
| 129 | Data integration | K | | | | | | 27 | | 27 |
| 130 | PNAS | H | | | | | | | 27 | 27 |
| 131 | Breast cancer | H | | | | | | | 27 | 27 |
| 132 | Neural networks | K | | | | | | 26 | | 26 |
| 133 | Interoperability | K | | | | | | 26 | | 26 |
| 134 | Knowledge management | K | | | | | | 26 | | 26 |
| 135 | Knowledge discovery | K | | | | | | 26 | | 26 |
| 136 | Friend | T | 26 | 4 | 22 | | | | | 26 |
| 137 | Report | T | 26 | 4 | 13 | 9 | | | | 26 |
| 138 | Age | T | 26 | 11 | 14 | | 1 | | | 26 |
| 139 | Startup | H | | | | | | | 26 | 26 |

| | Topics | Type | Term (T) | | | | | Keywords (K) | Hashtags (H) | Total |
|---|---|---|---|---|---|---|---|---|---|---|
| | | | **Total** | Blog | News | Policy | Wiki | | | |
| 140 | PMCON | H | | | | | | | 26 | 26 |
| 141 | Health informatics | H | | | | | | | 26 | 26 |
| 142 | Smart grid | K | | | | | | 25 | | 25 |
| 143 | Natural language processing | K | | | | | | 25 | | 25 |
| 144 | Proteomics | K | | | | | | 25 | | 25 |
| 145 | Network analysis | K | | | | | | 25 | | 25 |
| 146 | Online learning | K | | | | | | 25 | | 25 |
| 147 | HDFS | K | | | | | | 25 | | 25 |
| 148 | Nature | H | | | | | | | 25 | 25 |
| 149 | Microbiome | H | | | | | | | 25 | 25 |
| 150 | Publication | T | 25 | 8 | 1 | 12 | 4 | | | 25 |
| 151 | Patient | T | 25 | 3 | 20 | | 2 | | | 25 |
| 152 | High performance computing | K | | | | | | 24 | | 24 |
| 153 | Data management | K | | | | | | 24 | | 24 |
| 154 | Google | T | 24 | 4 | 19 | | 1 | | | 24 |
| 155 | Information | T | 24 | 2 | 9 | 4 | 9 | | | 24 |
| 156 | Landsat | H | | | | | | | 24 | 24 |
| 157 | Reliability | K | | | | | | 23 | | 23 |
| 158 | Parallel processing | K | | | | | | 23 | | 23 |
| 159 | Measurement | K | | | | | | 23 | | 23 |
| 160 | Distributed systems | K | | | | | | 23 | | 23 |
| 161 | Extreme learning machine | K | | | | | | 23 | | 23 |
| 162 | Information technology | K | | | | | | 23 | | 23 |
| 163 | Year | T | 23 | 9 | 13 | | 1 | | | 23 |
| 164 | Briefing | T | 23 | | 23 | | | | | 23 |
| 165 | Week | T | 23 | 12 | 11 | | | | | 23 |
| 166 | Time | T | 23 | 8 | 13 | | 2 | | | 23 |
| 167 | Bd2k | H | | | | | | | 23 | 23 |
| 168 | Support vector machine | K | | | | | | 22 | | 22 |
| 169 | Modeling | K | | | | | | 22 | | 22 |
| 170 | Neuroimaging | K | | | | | | 22 | | 22 |
| 171 | Data collection | K | | | | | | 22 | | 22 |
| 172 | Problem | T | 22 | 5 | 12 | | 5 | | | 22 |
| 173 | BMC bioinformatics | H | | | | | | | 22 | 22 |
| 174 | Bitcoin | H | | | | | | | 22 | 22 |
| 175 | Resource allocation | K | | | | | | 21 | | 21 |
| 176 | Supply chain management | K | | | | | | 21 | | 21 |
| 177 | GPU | K | | | | | | 21 | | 21 |
| 178 | Life | T | 21 | 6 | 14 | | 1 | | | 21 |

| | Topics | Type | Term (T) | | | | | Keywords (K) | Hashtags (H) | Total |
|---|---|---|---|---|---|---|---|---|---|---|
| | | | **Total** | Blog | News | Policy | Wiki | | | |
| 179 | Review | T | 21 | 15 | 5 | 1 | | | | 21 |
| 180 | Search | T | 21 | 4 | 17 | | | | | 21 |
| 181 | Health tech | H | | | | | | | 21 | 21 |
| 182 | Biobanks | H | | | | | | | 21 | 21 |
| 183 | Data visualization | H | | | | | | | 21 | 21 |
| 184 | Clinical | H | | | | | | | 20 | 20 |
| 185 | Smart manufacturing | H | | | | | | | 20 | 20 |
| 186 | Bioethics | H | | | | | | | 20 | 20 |
| 187 | Biodiversity | H | | | | | | | 20 | 20 |
| 188 | Issue | T | 20 | 14 | 5 | 1 | | | | 20 |
| 189 | Challenge | T | 20 | 8 | 12 | | | | | 20 |
| 190 | Heart disease | T | 20 | 4 | 16 | | | | | 20 |
| 191 | Tweet | T | 20 | 3 | 17 | | | | | 20 |
| 192 | Machine | T | 19 | 3 | 15 | | 1 | | | 19 |
| 193 | Liberal | T | 19 | 5 | 14 | | | | | 19 |
| 194 | GIS | H | | | | | | | 19 | 19 |
| 195 | 10 simple rules | H | | | | | | | 19 | 19 |
| 196 | Crowd sourcing | H | | | | | | | 19 | 19 |
| 197 | Citizen science | H | | | | | | | 19 | 19 |
| 198 | Part | T | 18 | 11 | 4 | | 3 | | | 18 |
| 199 | Conservative | T | 18 | 5 | 13 | | | | | 18 |
| 200 | Family | T | 17 | 6 | 11 | | | | | 17 |
| 201 | Link | T | 17 | 4 | 13 | | | | | 17 |
| 202 | Study find | T | 17 | 2 | 15 | | | | | 17 |
| 203 | App | T | 16 | 3 | 13 | | | | | 16 |
| 204 | State | T | 16 | 6 | 7 | 3 | | | | 16 |
| 205 | Development | T | 16 | 2 | 10 | 2 | 2 | | | 16 |
| 206 | Emotional contagion | T | 15 | 6 | 8 | | 1 | | | 15 |
| 207 | Instagram photo | T | 14 | 2 | 12 | | | | | 14 |
| 208 | Instagram post | T | 14 | | 14 | | | | | 14 |
| 209 | Woman | T | 14 | 3 | 11 | | | | | 14 |
| 210 | Facebook study | T | 14 | 7 | 7 | | | | | 14 |
| 211 | Future | T | 14 | 5 | 6 | 2 | 1 | | | 14 |
| 212 | Mood | T | 13 | 2 | 10 | | 1 | | | 13 |
| 213 | Users emotion | T | 13 | | 13 | | | | | 13 |
| 214 | Role | T | 13 | 5 | 7 | | 1 | | | 13 |
| 215 | Value | T | 13 | 4 | 5 | 3 | 1 | | | 13 |
| 216 | Web | T | 13 | 7 | 4 | | 2 | | | 13 |
| 217 | EU law | T | 12 | | | 12 | | | | 12 |
| 218 | Good cholesterol | T | 12 | | 12 | | | | | 12 |
| 219 | Sign | T | 12 | 1 | 11 | | | | | 12 |
| 220 | Expert | T | 12 | 1 | 10 | | 1 | | | 12 |
| 221 | Ethic | T | 11 | 6 | 3 | 2 | | | | 11 |
| 222 | Human | T | 11 | 5 | 4 | | 2 | | | 11 |

|  | Topics | Type | Term (T) |  |  |  |  | Keywords (K) | Hashtags (H) | Total |
|---|---|---|---|---|---|---|---|---|---|---|
|  |  |  | Total | Blog | News | Policy | Wiki |  |  |  |
| 223 | Need | T | 11 | 3 | 6 |  | 2 |  |  | 11 |
| 224 | Press release | T | 11 |  | 11 |  |  |  |  | 11 |
| 225 | Term | T | 11 | 1 | 2 |  | 8 |  |  | 11 |
| 226 | Opportunity | T | 11 | 5 | 6 |  |  |  |  | 11 |
| 227 | University | T | 11 | 2 | 2 |  | 7 |  |  | 11 |
| 228 | Facebook like | T | 10 | 3 | 7 |  |  |  |  | 10 |
| 229 | Wikipedia | T | 10 | 4 | 5 |  | 1 |  |  | 10 |
| 230 | Poverty | T | 9 |  | 8 | 1 |  |  |  | 9 |
| 231 | Big data approach | T | 9 | 1 | 8 |  |  |  |  | 9 |
| 232 | Lesson | T | 9 | 7 | 2 |  |  |  |  | 9 |
| 233 | Doctor | T | 9 | 2 | 7 |  |  |  |  | 9 |
| 234 | Instagram photos | T | 9 | 1 | 8 |  |  |  |  | 9 |
| 235 | Nanotechnology | T | 9 |  | 9 |  |  |  |  | 9 |

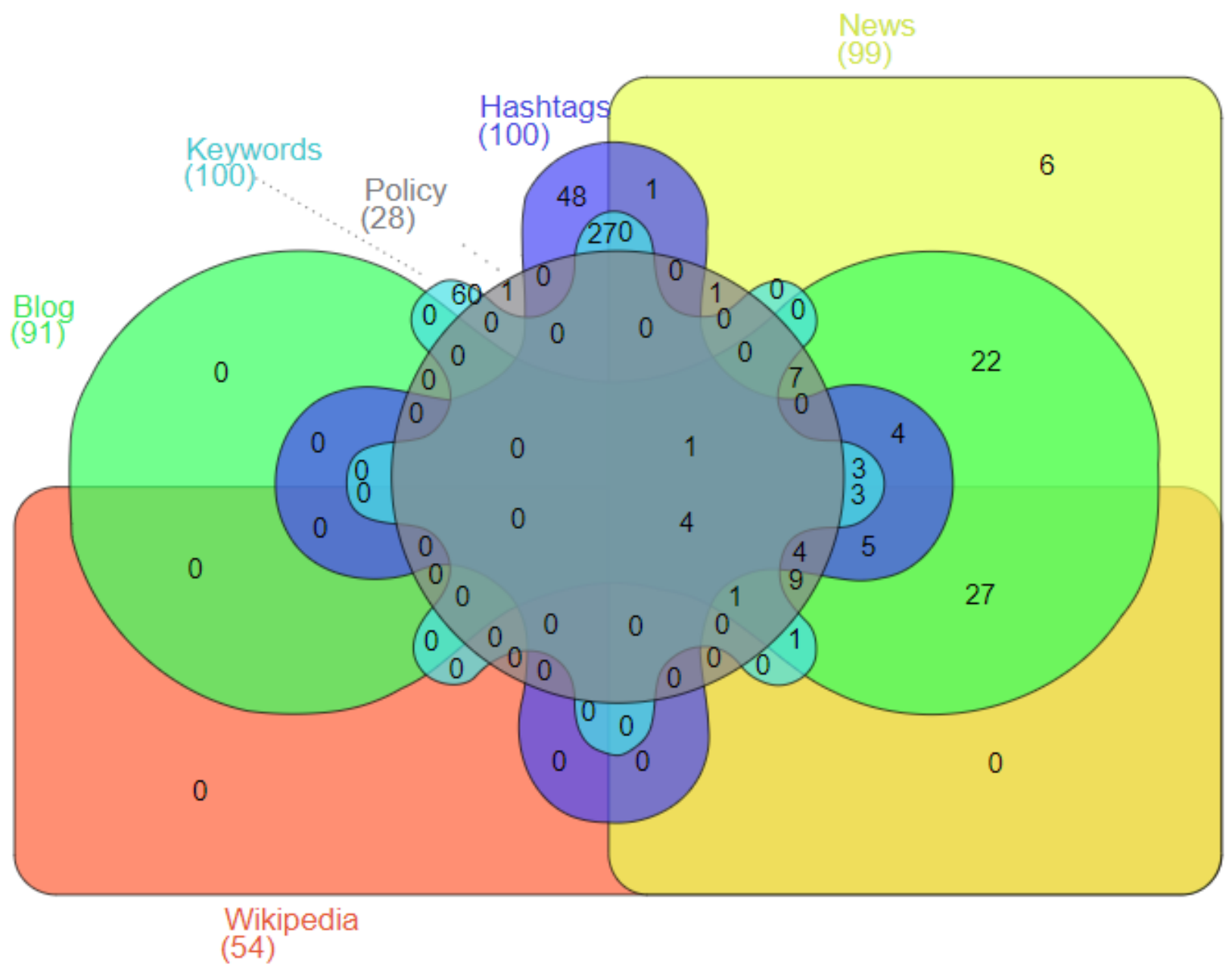

**Fig. 14** Venn diagram of topic sets in six groups

**Table 9** Cosine similarity of topic sets in six groups

|  | Blog | Policy | News | Wikipedia | Papers | Twitter |
|---|---|---|---|---|---|---|
| **Blog** |  | 0.5151 | 0.9587 | 0.7703 | 0.1363 | 0.2516 |
| **Policy** |  |  | 0.5128 | 0.4629 | 0.1134 | 0.1701 |
| **News** |  |  |  | 0.7385 | 0.1307 | 0.2513 |
| **Wikipedia** |  |  |  |  | 0.1225 | 0.2177 |
| **Papers** |  |  |  |  |  | 0.3900 |
| **Twitter** |  |  |  |  |  |  |